\documentclass[usenatbib,usegraphicx]{mn2e}
\usepackage{amssymb}
\usepackage{bm}
\usepackage{color}
\usepackage{graphicx}

\input epsf

\begin{document}

\def\Lya{Ly$\alpha$}
\def\Lyb{Ly$\beta$}
\def\DI{\hbox{D~$\scriptstyle\rm I\ $}}
\def\HI{\hbox{H~$\scriptstyle\rm I\ $}}
\def\bHI{\hbox{\bf H~$\scriptstyle\bf I\ $}}
\def\HII{\hbox{H~$\scriptstyle\rm II\ $}}
\def\HeI{\hbox{He~$\scriptstyle\rm I\ $}}
\def\HeII{\hbox{He~$\scriptstyle\rm II\ $}}
\def\HeIII{\hbox{He~$\scriptstyle\rm III\ $}}
\def\OIII{\hbox{O~$\scriptstyle\rm III\ $}}
\def\kms{\,{\rm {km\, s^{-1}}}}
\def\msun{{M_\odot}}
\def\nHI{{\rm HI}}
\def\nH{{\rm H}}
\def\nHII{{\rm HII}}
\def\mpc{\,{\rm {Mpc}}}
\def\kmsmpc{\,{\rm km\,s^{-1}\,Mpc^{-1}}}
\def\hmpc{\,h^{-1}{\rm \,Mpc}\,}
\def\mpch{\,h{\rm \,Mpc}^{-1}\,}
\def\hkpc{\,h^{-1}{\rm \,kpc}\,}
\def\ev{\,{\rm eV\ }}
\def\kel{\,{\rm K\ }}
\def\intunits{\,{\rm ergs\,s^{-1}\,cm^{-2}\,Hz^{-1}\,sr^{-1}}}
\def\ltsima{$\; \buildrel < \over \sim \;$}
\def\lsim{\lower.5ex\hbox{\ltsima}}
\def\gtsima{$\; \buildrel > \over \sim \;$}
\def\gsim{\lower.5ex\hbox{\gtsima}}
\def\etal{{ et~al.~}}
\def\aap{A\&Ap}
\def\aj{AJ}
\def\ana{A\&A}
\def\apj{ApJ}
\def\apjl{ApJ}
\def\apjs{ApJS}
\def\apss{ApSS}
\def\araa{ARA\&A}
\def\bain{Bull. Ast. Insts. Neth.}
\def\jcap{JCAP}
\def\mnras{MNRAS}
\def\nat{Nature}
\def\prd{PhysRevD}
\def\spose#1{\hbox to 0pt{#1\hss}}
\def\lta{\mathrel{\spose{\lower 3pt\hbox{$\mathchar"218$}}
     \raise 2.0pt\hbox{$\mathchar"13C$}}}
\def\gta{\mathrel{\spose{\lower 3pt\hbox{$\mathchar"218$}}
     \raise 2.0pt\hbox{$\mathchar"13E$}}}

\journal{Preprint-00}

\title{The impact of \Lya\ emission line heating and cooling on the cosmic dawn 21-cm signal}

\author[Avery Meiksin, Piero Madau]{ Avery
  Meiksin$^{1}$\thanks{E-mail:\ meiksin@ed.ac.uk (AM)}, Piero
  Madau$^{2, 3}$\\
  $^{1}$SUPA\thanks{Scottish Universities Physics Alliance}, Institute
  for Astronomy, University of Edinburgh,
  Blackford Hill, Edinburgh\ EH9\ 3HJ, UK\\
  $^{2}$Department of Astronomy and Astrophysics, University of
  California, Santa Cruz, 1156 High Street, Santa Cruz, CA~95064, USA\\
  $^{3}$Institute of Astronomy, University of Cambridge, Madingley
  Road, Cambridge CB3~0HA, UK}


\maketitle

\begin{abstract}
  Allowing for enhanced \Lya\ photon line emission from Population~III
  dominated stellar systems in the first forming galaxies, we show the
  21-cm cosmic dawn signal at $10<z<30$ may substantially differ from
  standard scenarios. Energy transfer by \Lya\ photons emerging from
  galaxies may heat intergalactic gas if \HII\ regions within galaxies
  are recombination bound, or cool the gas faster than by adiabatic
  expansion if reddened by winds internal to the haloes. In some
  cases, differential 21-cm antenna temperatures near $-500$~mK may be
  achieved at $15<z<25$, similar to the signature detected by the
  EDGES 21-cm cosmic dawn experiment.
\end{abstract}


\section{Introduction}

The first bound stellar systems to form in the Universe have long been
believed to be copious emitters of \Lya\ photons
\citep{1967ApJ...147..868P}, possibly in a bright flash brought on by
a violent burst of star formation as the first massive stars formed,
ionized their surroundings and subsequently exploded as supernovae
\citep{1964ApJ...140.1434F, 1971ApSS..14..179T,
  1972ApJ...178..319T}. The first generation of stars will form
through gas processes, and so have very low metallicity. These
Population~III dominated stellar populations are expected to have an
excess of massive stars radiating strongly in the Lyman
continuum. Larger \Lya\ equivalent widths are predicted compared with
current galaxies, with values as high as a few thousand Angstr\"oms
possible \citep{2010AA...523A..64R}. While the fraction of
star-forming galaxies with strong \Lya\ emission is observed to
decrease at $z>6$ \citep[eg][]{2010MNRAS.408.1628S,
  2012ApJ...744...83O, 2014ApJ...795...20S}, this is often interpreted
as a consequence of the resonance line scattering of \Lya\ photons in
an increasingly netural intervening intergalactic medium (IGM) at
early epochs \citep[eg][]{2011MNRAS.414.2139D}.

There is scant direct observational data on star-forming galaxies at
$z>8$. The global rate of \Lya\ photon production from these systems
is consequently unknown, as well as the halo masses in which they may
reside. Whilst haloes with masses as low as $\sim10^6\,h^{-1}M_\odot$
may achieve the densities required to form sufficient molecular
hydrogen to cool the halo gas and drive star formation, the UV
radiation from the first galaxies at $z>20$ may dissociate the
molecular hydrogen in subsequent generations, suppressing further star
formation \citep{1997ApJ...476..458H}. Star formation would then be
restricted to haloes more massive than $\sim10^7\,h^{-1}M_\odot$ that
cool through atomic processes. How effective molecular hydrogen
dissociation is as a means of suppressing star formation in less
massive haloes is currently unknown.

Whilst direct observations of \Lya-emitting galaxies at $z>20$ are
likely someway off, such galaxies may have a detectable impact on the
21-cm signature expected from the still neutral IGM
\citep{1997ApJ...475..429M}. Under conventional assumptions of star
formation in primordial galaxies, with a minimum halo mass of
$\sim10^7\,h^{-1}M_\odot$, the metagalactic \Lya\ photon scattering
rate $P_\alpha$ arising from \HII\ regions within the galaxies at
$z>25$ is typically below the thermalization rate $P_{\rm th}$
required to drive the 21-cm spin temperature to the kinetic
temperature of the IGM. Strong coupling to the kinetic temperature,
with $P_\alpha>10P_{\rm th}$, would not be achieved until $z<20$. As a
consequence, only a weak absorption signal against the CMB is expected
at $z>20$, with $\vert\Delta T_{\rm 21-cm}\vert<100$~mK
\citep{2005ApJ...626....1B, 2006MNRAS.367..259H}.

In this paper, we show that, allowing for enhanced \Lya\ photon
production in Pop~III star \HII\ regions and extending the minimum
halo mass for star formation to include molecular hydrogen cooled
haloes, the metagalactic \Lya\ photon background may be boosted at
$z>20$ to a level that substantially alters the predicted strength of
the 21-cm signature.

We also show that for extreme cases of \Lya\ photon emission, with
equivalent widths exceeding $10^5$~\AA\ or for haloes emitting to the
red of line-centre with moderate equivalent widths but narrow line
widths, it may be possible to cool the IGM super-adiabatically, with a
21-cm differential temperature $\Delta T_{\rm 21-cm}<-400$~mK at
$15<z<26$, similar to the recent signal reported for the EDGES global
21-cm signature experiment \citep{2018Natur.555...67B}. Whether such
extreme conditions may be realized in astrophysical sources, however,
is unknown.

This paper is organized as follows. In the next section, the possible
level of \Lya\ emission from primordial galaxies is assessed. In
Sec.~III, we describe the method for solving for the metagalactic
\Lya\ radiation field from the sources. Solutions for the resulting
evolution of the IGM kinetic temperature are presented in Sec.~IV,
allowing for \Lya\ photon recoil heating and cooling of the IGM. The
results are discussed in Sec.~V, and we summarize our conclusions in
Sec.~VI. We adopt a flat $\Lambda$CDM cosmological model with
$\Omega_m=0.3111$, $\Omega_bh^2=0.02242$, $h=0.6766$,
$\sigma_8=0.8102$ and spectral index $n=0.9665$
\citep{2018arXiv180706209P}. A comoving kiloparsec is designated \lq
ckpc.\rq

\section{\Lya\ photon sources}
\label{sec:sources}

\subsection{Emergent \Lya\ emission from primordial
  star-forming galaxies}
\label{subsec:haloes}

We consider three varieties of \Lya\ photons produced by star-forming
galaxies:\ (a)\ directly produced \Lya\ photons from \HII\ regions
within the galaxies, (b)\ continuum photons which redshift to the
local \Lya\ resonance frequency as a result of cosmological expansion,
and (c)\ continuum photons which redshift to higher order Lyman
transitions that produce \Lya\ photons through radiative
cascades. Averaged over cosmological scales, the first type dominates,
as we shall discuss in this section. We begin by characterising the
emission line profiles emergent from the galaxies.

In a metal-free environment, a minimum halo mass between
$10^5-10^7\,M_\odot$ is required to form stars at $z>15$ in a Cold
Dark Matter dominated cosmology following molecular hydrogen
production through gaseous processes and cooling
\citep{1983ApJ...274..443B,1984Natur.311..517B, 1986MNRAS.221...53C,
  2003ApJ...592..645Y}. Because of the relative streaming velocity
between baryons and dark matter at the recombination era, however, the
gas content of lower mass haloes is suppressed
\citep{2010PhRvD..82h3520T}. Although estimates vary, the lowest mass
haloes allowing substantial star formation likely exceeds
$\sim10^6\,h^{-1}M_\odot$ \citep[eg][]{2012Natur.487...70V,
  2019MNRAS.484.3510S}.

The typical central hydrogen density and
temperature of a $10^6\,M_\odot$ halo collapsing at $z=20$ are
$n_{\rm H}\simeq10\,{\rm cm^{-3}}$ and $T\simeq10^3$~K,
respectively, within a core comoving radius $r_c\simeq3$~ckpc
\citep[eg][]{2011MNRAS.417.1480M}.
The formation of \HII\ regions around the more massive stars will
produce \Lya\ photons. If dust absorption is low, the \Lya\ photons
will scatter until they escape the galaxy. Because the interstellar
medium (ISM) of the galaxies will typically be highly optically thick
to \Lya\ photons, the emitted profile will be double-peaked about the
line centre, as the \Lya\ photons scatter in frequency until they are
sufficiently removed from line centre to escape
\citep{1949BAN....11....1Z, 1973MNRAS.162...43H}.

An exception to the double-peaked profile may arise if the ISM is
expanding, as in a wind powered by hot stars or supernovae. In this
case the profile may be highly reddened. Such reddened \Lya\ profiles
are common in star-forming galaxies \citep[eg][]{2010ApJ...717..289S,
  2011ApJ...730....5H}. Numerical simulations suggest the halo gas may
become highly rarefied following expulsion of most of the gas by
supernovae, and cooling may lower the gas temperature to below 1000~K
\citep{2008ApJ...682...49W, 2013ApJ...774...64W,
  2015MNRAS.452.2822S}. In such circumstances, the reddened profile
may peak only several Doppler widths to the red of line centre. We
shall see below such sources can cool the IGM if their combined
intensity is sufficiently strong. Both double-peaked and reddened
emergent \Lya\ profiles are considered.

\subsection{Metagalactic \Lya\ photon scattering rate}
\label{subsec:scattering}

We quantify the \Lya\ emission intensity from the galaxies in terms of
the metagalactic \Lya\ scattering rate per hydrogen atom they
produce. For a specific number density $n_\nu$ of \Lya\ photons, the
\Lya\ photon scattering rate per atom in the lower state is
\begin{equation} 
P_\alpha=\sigma_\alpha c\int_0^\infty\,d\nu \varphi_V(\nu) n_\nu,
\label{eq:Palpha}
\end{equation} 
where the total scattering cross section is
$\sigma_\alpha=(\pi e^2/m_ec)f_\alpha$ with upper oscillator strength
$f_\alpha$ and $\varphi(\nu)$ is the Voigt scattering profile.
The characteristic scattering
rate for which \Lya\ photons couple the spin temperature of the 21-cm
hyperfine transition to the light temperature instead of to the CMB
temperature through the Wouthuysen-Field mechanism
\citep{1952AJ.....57R..31W, 1958PROCIRE.46..240F} is given by the
thermalization scattering rate
\begin{equation} 
P_{\rm th}=\frac{27A_{10}T_{\rm  
    CMB}(z)}{4T_*}\simeq(1.6\times10^{-11}\,{\rm  
  s}^{-1})\left(\frac{1+z}{21}\right),
\end{equation}
\citep{1997ApJ...475..429M}. Here, $A_{10}$ is the spontaneous decay
rate of the 21-cm transition and $T_*=h\nu_{10}/k$ is the temperature
corresponding to the 21-cm transition of frequency $\nu_{10}$. For
$P_\alpha>>P_{\rm th}$, the spin temperature will track the light
temperature, which will be close to the kinetic temperature of the IGM
\citep{1959ApJ...129..536F, 2006MNRAS.370.2025M}.

\begin{figure}
\scalebox{0.47}{\includegraphics{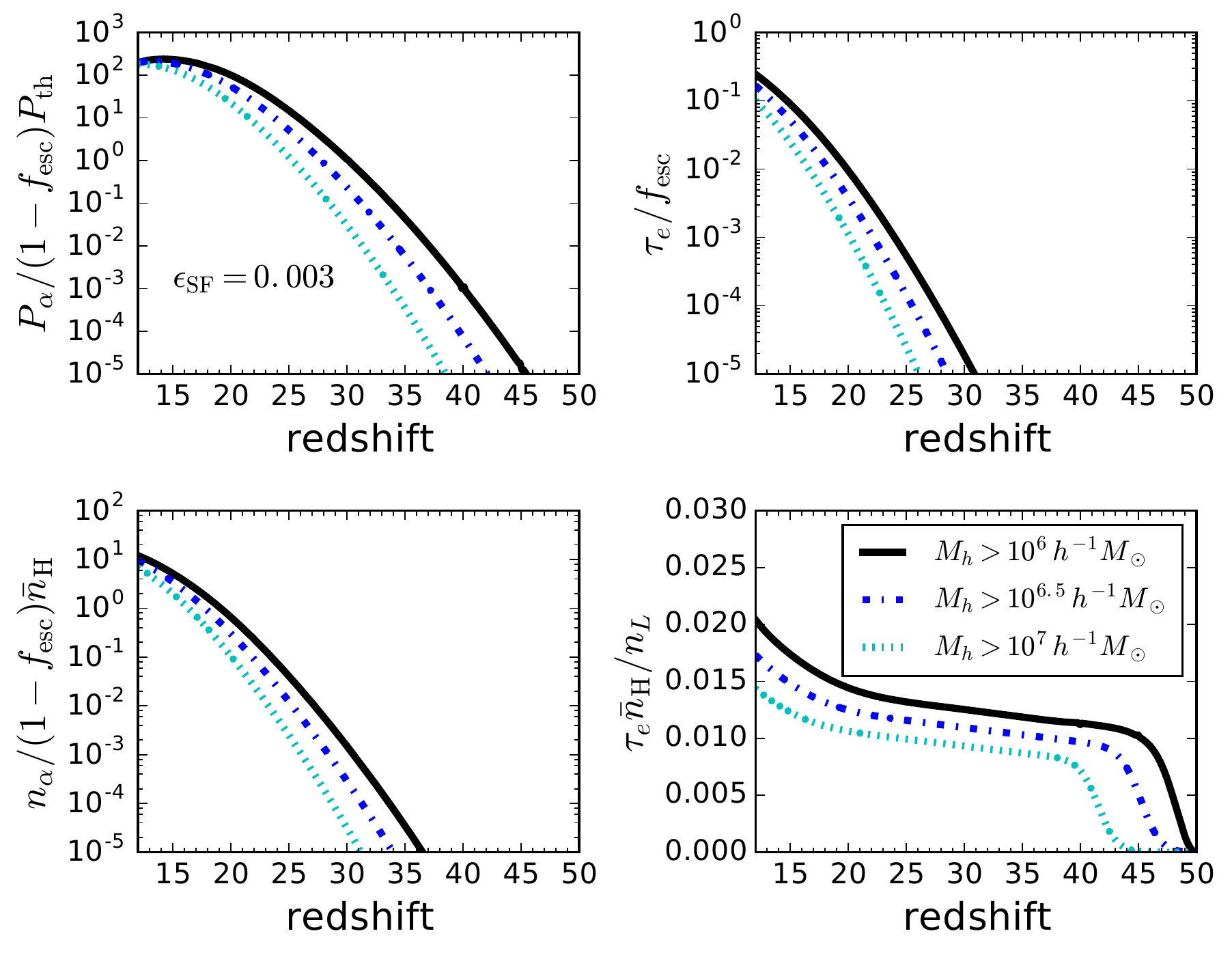}}
\caption{Evolution of the \Lya\ scattering rate $P_\alpha$, normalized
  by the thermalization rate $P_{\rm th}$ (top left panel), integrated
  number density $n_\alpha$ of \Lya\ photons produced, normalized by
  the mean IGM hydrogen density $\bar n_{\rm H}$ (bottom left panel),
  evolution of the Thomson optical depth (top right panel), and of the
  Thomson optical depth scaled to $n_L/\bar n_{\rm H}$, where $n_L$ is
  the number of Lyman continuum photons produced in the IGM for
  $f_{\rm esc}>0$ (bottom
  right panel). A fiducial baryon conversion efficiency of
  $\epsilon_{\rm SF}=0.003$ is assumed. The curves correspond to the
  indicated minimum halo mass cut-off values for star formation.
}
\label{fig:Pana_evol}
\end{figure}

To maintain a scattering rate $P_\alpha$ in an IGM with a Sobolev
parameter $\gamma_S=H(z)\nu_0/(\bar n_{\rm H}\sigma_\alpha c)$, where
$\nu_0$ is the \Lya\ photon resonance frequency, $H(z)$ is the Hubble
parameter and $\bar n_{\rm H}$ the average intergalactic hydrogen
density, \Lya\ photons must be generated at the rate
\begin{equation}
  \dot n_\alpha = \gamma_S \bar n_{\rm H}P_\alpha
  \label{eq:nadot}
\end{equation}
\citep{1959ApJ...129..536F, 2010MNRAS.402.1780M}. For a net efficiency
$\epsilon_{\rm SF}$ for converting gaseous baryons in haloes into
nuclear fuel within stars, allowing
$L_{\rm Ly\alpha}=f_{\rm Ly\alpha}L_{\rm bol}$ where $L_{\rm bol}$ is
the stellar bolometric luminosity, with
$f_{\rm Ly\alpha}\simeq0.2-0.4$ for low metallicity (Pop~III) stars with a
Salpeter stellar initial mass function \citep{2010AA...523A..64R}, and
an energy generation rate per unit mass $0.007c^2$ for nuclear
burning, the total generation rate of \Lya\ photons integrated over
halo masses is:
\begin{equation}
  \dot n_\alpha = (0.001-0.003)\frac{\epsilon_{\rm SF}}{e_\alpha}\frac{\Omega_b}{\Omega_m}\frac{d}{dt}\int_{M_{\rm min}} dM_h \frac{dn_h(M_h)}{dM_h}M_hc^2,
\end{equation}
for a differential number density of halos $dn_h(M_h)/dM_h$, minimum
halo mass $M_{\rm min}$ for forming stars, and \Lya\ photon energy
$e_\alpha$. Typically, $\epsilon_{\rm SF}\simeq0.001-0.01$.
\footnote{Setting $\epsilon_{\rm SF}=f_*/3$, where $f_*$ is the
  conversion efficiency of gas mass in haloes to stellar mass, and
  adopting a Pop~II value of $f_{\rm Ly\alpha}=0.03$ and nebular \Lya\
  equivalent width of 200~\AA \citep{2010AA...523A..64R}, matches the
  number of photons produced per stellar baryon by the stellar
  continuum between Ly$\alpha$\ and Ly$\beta$ of $\sim6500$ given by
  \citet{2005ApJ...626....1B}.} The corresponding \Lya\ equivalent
widths range over $400-2000$~\AA\ for continuous star formation and
$2000-4000$~\AA\ for a young burst population
\citep{2010AA...523A..64R}, with the higher values corresponding to an
extended upper stellar mass of $500M_\odot$, and truncating the lower
end at $50M_\odot$ for the highest values.

The \Lya\ scattering rate evolves rapidly, as shown in
Fig.~\ref{fig:Pana_evol}. Depending on the minimum halo mass for
producing stars, by $z\approx25-35$, the scattering rate exceeds the
thermalization rate and the \Lya\ photons are able to couple the spin
temperature to the gas temperature. The cumulative number density of
\Lya\ photons produced rises rapidly as well, because of the rapid
evolution in the number of haloes above the threshold mass,
potentially exceeding the mean number of intergalactic hydrogen atoms
by $z\approx20$. Such a high number is problematic in terms of
reionization of the IGM, because the production
rate of Lyman continuum photons from the stars is comparable to the
production rate of \Lya\ photons in \HII\ regions. A \Lya\ photon flux
sufficient to provide moderate coupling of the IGM spin temperature to the
kinetic temperature requires an escape fraction $f_{\rm esc}\lta0.2$ of
ionizing radiation from the galaxies \citep{2018MNRAS.480L..43M}.

Details of the relation between the \Lya\ scattering rate and
reionization limits are discussed in the Appendix. There it is shown
that, for an escape fraction $f_{\rm esc}<0.2$, requiring
$n_L/\bar n_{\rm H}<0.5$ imposes the restriction
$P_\alpha/P_{\rm th}\lta200$ for $\epsilon_{\rm SF}=0.003$. Even for
the extreme efficiency $\epsilon_{\rm SF}=0.1$,
$P_\alpha/P_{\rm th}<3000$ is required, while for
$\epsilon_{\rm SF}=10^{-4}$, $P_\alpha/P_{\rm th}<20$ is required. The
limits become even more severe for higher escape fractions.

\subsection{\Lya\ photon heating of the IGM}
\label{subsec:heating}

The required metagalactic \Lya\ photon scattering rate to alter the
thermal evolution of the IGM may be quantified as follows. For a \Lya\
photon number density $n_\nu$, the volumetric heating rate of the IGM
resulting from atomic recoils is
\begin{equation} 
G_{\rm H}=P_\alpha n_{\rm H} \frac{h\nu_0}{m_{\rm H} c^2} h\nu_0 \left(1-\frac{T_K}{T_L}\right),
\label{eq:GheatVTn}
\end{equation} 
\citep{2006MNRAS.370.2025M}, where $m_{\rm H}$ is the mass of a
hydrogen atom, $T_K$ is the IGM kinetic temperature and $T_L$ is the
light temperature of the \Lya\ radiation field (see below). The limit
$T_K << T_L$ of Eq.~(\ref{eq:GheatVTn}) recovers the heating rate
given by \citet{1997ApJ...475..429M}. At lower light temperatures, the
heating rate is modified by the factor $1-T_K/T_L$, which could result
either in net heating or cooling \citep[cf][]{2004ApJ...602....1C,
  2006ApJ...651....1C}.

To change the gas temperature within a Hubble time requires a heating
rate $G_{\rm H}(2/3H) > (3/2)n_{\rm H}k_{\rm B}T_K$, or a
scattering rate
\begin{equation}
    P_\alpha > \frac{9}{8}H(z)\epsilon^{-2}\left(1-\frac{T_K}{T_L}\right)^{-1},
    \label{eq:Palim}
\end{equation}
where $\epsilon=h\nu_0/(2kT_Km_{\rm H}c^2)^{1/2}\simeq0.025T_K^{-1/2}$
is the recoil parameter. In a standard cosmological expansion
scenario, the diffuse IGM temperature at $z=20$ is $T_K\simeq10$~K
\citep{2000ApJS..128..407S}. We shall show that typically
$\vert 1-T_K/T_L\vert\sim10^{-4}-10^{-3}$.  Then the required \Lya\
scattering rate is
\begin{eqnarray}
P_\alpha &>& (2.2\times10^{-9}\,{\rm s}^{-1})\frac{T_K}{10~{\rm
    K}}\left\vert(10^3)\left( 1-\frac{T_K}{T_L}\right)\right\vert^{-1}\nonumber\\
& =& (10^2-10^3)P_{\rm th}.
\label{eq:Pamin}
\end{eqnarray}

Additional \Lya\ photons are provided by cascades from higher order
Lyman photon scattering produced by continuum source photons
redshifting into higher order Lyman resonances
\citep{2005ApJ...626....1B}. Higher order Lyman photons scattering
within the Doppler core will also heat the gas near the sources. The
cascade-produced \Lya\ photons and higher order Lyman photon heating
are accounted for following \citet{2010MNRAS.402.1780M}. Details are
provided in an Appendix.

\section{The metagalactic \Lya\ radiation background}
\label{sec:Lyabg}

The heating rate by \Lya\ photons is sensitive to the shape of the
metagalactic \Lya\ radiation background. We solve for the frequency
dependence of the radiation field in the diffusion approximation. We
shall consider both directly emitted \Lya\ photons from galaxies and
continuum photons that redshift to the local \Lya\ frequency. At
$z=20$, the comoving density of haloes more massive than
$10^6h^{-1}M_\odot$ is $\sim140\,{\rm cMpc}^{-3}$
\citep{2007MNRAS.374....2R}, corresponding to a mean comoving
separation of $\lta200$~ckpc. For comparison, the mean-free-path of
\Lya\ photons in the IGM even up to 100 Doppler widths from line
centre is much smaller, $\sim15$~ckpc, so that the \Lya\ radiation
field surrounding the galaxies may be treated in the diffusion
approximation for both directly emitted \Lya\ photons and those
arising from the redshifted galaxy continua. The photons diffuse
through the IGM on scales up to $\sim1$~Mpc (proper, independent of
redshift) before free-streaming becomes important
\citep{1999ApJ...524..527L, 2012MNRAS.426.2380H}. This corresponds to
a region containing $10^6-10^7$ haloes more massive than
$10^6\,M_\odot$. The large number of galaxies within the diffusive
region of the \Lya\ photons ensures a smoothly varying radiation field
throughout the IGM. Accordingly, the galactic sources of \Lya\ photons
will be treated as isotropic and homogeneously spatially distributed.

In a homogeneous and isotropic expanding medium, the radiative
transfer equation for the specific comoving number density $n(\nu,t)$
of \Lya\ photons of frequency $\nu$ at time $t$ produced by a comoving
source density $S(\nu,t)$ is solved in the diffusion approximation for
a homogeneous and isotropic radiation field. Defining the
dimensionless frequency shift $x=(\nu-\nu_0)/\Delta\nu_D$, where
$\nu_0$ is the \Lya\ resonance transition frequency,
$\Delta\nu_D=\nu_0(2kT/m_{\rm H})^{1/2}/c$ is the Doppler width, a
conformal time
$\tau=\int_0^t dt n_{\rm H}(t)\sigma_\alpha c/\Delta\nu_D$ for neutral
hydrogen density $n_{\rm H}$, and the dimensionless Voigt profile
$\phi_V(x)=\varphi_V(\nu)\Delta\nu_D$, the diffusion equation takes
the dimensionless form
\begin{eqnarray}
\frac{\partial n_x(\tau)}{\partial\tau}&-&\frac{\partial\log\Delta\nu_D}{\partial\tau}n_x(\tau)\nonumber\\
&=&\frac{1}{2}\frac{\partial}{\partial x}\left\{2\left[\epsilon\phi_V(x)+\gamma_S\right]n_x(\tau) +
    \phi_V(x)\frac{\partial}{\partial x}n_x(\tau)\right\}\nonumber\\
&+&\tilde S(x,\tau),
\label{eq:nxevFP}
\end{eqnarray}
\citep{1994ApJ...427..603R, 2004ApJ...602....1C, 2006MNRAS.370.2025M},
where $n_x(\tau)=n_\nu(t)\Delta\nu_D$ and
$\tilde S(x,\tau)=S(\nu)(\Delta\nu_D)^2/(n_{\rm H}\sigma_\alpha c)$
have been defined, a possible time-evolution of $\Delta\nu_D$ has been
allowed for and terms of order $xb/c$ have been neglected. The recoil
parameter $\epsilon$
accounts for
atomic recoil after collision with a \Lya\ photon. The valid recovery
of the effects of atomic recoil at this level of approximation
requires $b/c << 2\epsilon$, or $T<<h\nu_0/k\approx10^5$~K for
hydrogen \Lya\ photons.

At low temperatures ($T<4$~K), corrections to the radiative transfer
equation arising from the fine structure of the \Lya\ resonance and
spin-flip scatterings start to become relevant
\citep{2006MNRAS.367..259H}. These lead to corrections to the recoil
parameter and effective \Lya\ scattering rate. The correction factor
for the recoil parameter depends on the spin temperature, given by
\begin{equation}
T_S = \frac{T_{\rm CMB} + y_\alpha T_L + y_cT_K}{1+y_\alpha+y_c},
\label{eq:TSpin}
\end{equation}
where $y_\alpha$
and $y_c$
are weight factors depending on the Ly$\alpha$
and collisional de-excitation rates \citep{1959ApJ...129..536F,
  1997ApJ...475..429M, 2006MNRAS.370.2025M}. Since we shall
investigate the case of strong \Lya\ scattering, Eq.~(\ref{eq:TSpin})
shows the spin temperature goes over to the IGM kinetic temperature,
since the light temperature quickly converges to the kinetic
temperature \citep{2006MNRAS.370.2025M}, resulting in a negligible
change to the recoil parameter. In this limit, the \Lya\ scattering
rate may be adjusted by reducing the Sobolev parameter by the factor
$1+0.4/
T_K$ \citep{2006ApJ...651....1C}. The effect on the 21-cm differential
temperatures in the contexts we examine is very small.

The left hand side of Eq.~(\ref{eq:nxevFP}) is of order
$(H\lambda_{\rm mfp}/c)n_x$, much smaller than the right hand
side, except possibly far in the wings. Accordingly, the number
density of photons may be taken as time-steady in the comoving
frame. The diffusion approximation then becomes
\begin{eqnarray}
\phi_V(x)\frac{\partial}{\partial x}n_x(\tau) &+& 2\epsilon\phi_V(x)n_x(\tau) +
    2\gamma_S n_x(\tau)\nonumber\\
&=&-2\int_{-\infty}^x\,dx^\prime\,\tilde S(x^\prime,\tau) + 2\gamma_S n_{-\infty}(\tau),
\label{eq:nxevFP_ss}
\end{eqnarray}
where $2\gamma_S n_{-\infty}(\tau)$ is an integration constant. The
source function and boundary conditions on $n_x(\tau)$ satisfy the
integral constraint
\begin{equation}
\int_{-\infty}^\infty\,dx\,\tilde
S(x,\tau)=\gamma_S\left[n_{-\infty}(\tau)-n_\infty(\tau)\right]
\label{eq:Sxconstraint}
\end{equation}
given by the asymptotic values of the radiation field. In the absence
of a source, the asymptotic values correspond to a flat background
continuum, with $n_\infty=n_{-\infty}$.

Allowing for both a source and a background continuum is readily
accommodated by fixing $n_\infty$ to the continuum value, arising from
the continuum emission from galaxies. Because continuum photons
emitted between \Lya\ and \Lyb\ redshift to local \Lya\ photons, a
source that injects the same number of \Lya\ photons directly will
have an equivalent width of
$\lambda_\alpha-\lambda_\beta\simeq190$~\AA. Sources with \Lya\
equivalent widths of $1000-3000$~\AA\ will then produce 5--15 times as
many \Lya\ photons directly as produced by the continuum. Additional
\Lya\ photons may also be produced from higher energy continuum
photons that redshift to higher order Lyman resonances where they are
scattered and subsequently cascade to \Lya\ photons
\citep{2004ApJ...602....1C}.

The light temperature in Eq.~(\ref{eq:GheatVTn}) is given by
\begin{equation}
T_L=\int_0^\infty\, dx n_x \phi_V(x){\Bigg/}
\int_0^\infty\, dx n_x\phi_V(x)\frac{1}{T_n(x)},
\label{eq:TnH}
\end{equation} 
\citep{2006MNRAS.370.2025M}, where  
\begin{equation} 
T_n(x)=-\frac{h\Delta\nu_D}{k}\left(\frac{d\log n_x}{dx}\right)^{-1}
\label{eq:Tn}
\end{equation} 
\citep[cf][]{2006ApJ...647..709R}. Using Eq.~(\ref{eq:nxevFP_ss}), the
light temperature may be expressed in the often numerically more
practical alternative form
\begin{eqnarray}
T_L &=&
-\frac{h\Delta\nu_D}{k}\int_{-\infty}^\infty\,dx\,\phi_V(x)n_x\nonumber\\
&&\times\Biggl\{\int_{-\infty}^\infty\,dx\,\left[2\gamma_S\left(n_{-\infty}-n_x\right)-{\mathbf
      2}\int_{-\infty}^x\,dx^\prime\,S(x^\prime)\right]\nonumber\\
    &-&
    2\epsilon\int_{-\infty}^\infty\,dx\,\phi_V(x)n_x\Biggr\}^{-1}
\end{eqnarray}
\citep[cf][]{2004ApJ...602....1C}. In a static medium, $T_L$ rapidly
converges to $T_K$, resulting in no net heating or cooling
\citep{2006MNRAS.370.2025M}. By contrast, in a dynamical medium the
light temperature may never reach the kinetic temperature, resulting
in a non-vanishing heating or cooling rate
\citep{2004ApJ...602....1C}.

The kinetic temperature $T_K$ of the IGM
evolves according to
\begin{equation}
  \frac{dT_K}{dz} = -\frac{2}{1+z}T_K -
  \frac{2}{3}\frac{1}{(1+z)H(z)}\frac{G_{\rm H}}{nk},
\end{equation}
for particle density $n\simeq 1.1n_{\rm H}$, allowing for helium and
taking the IGM to be almost entirely neutral.

The measured differential 21-cm antenna temperature between the IGM spin
temperature at the mean IGM density and the CMB is
\begin{equation}
\delta T_{\rm 21-cm}=(1+z)^{-1}\left[T_S(z) - T_{\rm
    CMB}(z)\right]\left[1-e^{-\tau_{21}(z)}\right],
\label{eq:dT21cm}
\end{equation} 
where the intergalactic 21-cm optical depth is
\begin{equation}
\tau_{21}(z)\simeq0.402\frac{x_{\rm HI}(z)}{T_S(z)}\left(\frac{1+z}{13}\right)^{1.5},
\label{eq:tau21cm}
\end{equation}
\citep{1997ApJ...475..429M}, and $x_{\rm HI}(z)$ is the neutral
fraction at redshift $z$.

\section{Evolution of differential 21-cm antenna temperature}

\subsection{Static ISM models}
\label{subsec:static}

\begin{figure}
\scalebox{0.43}{\includegraphics{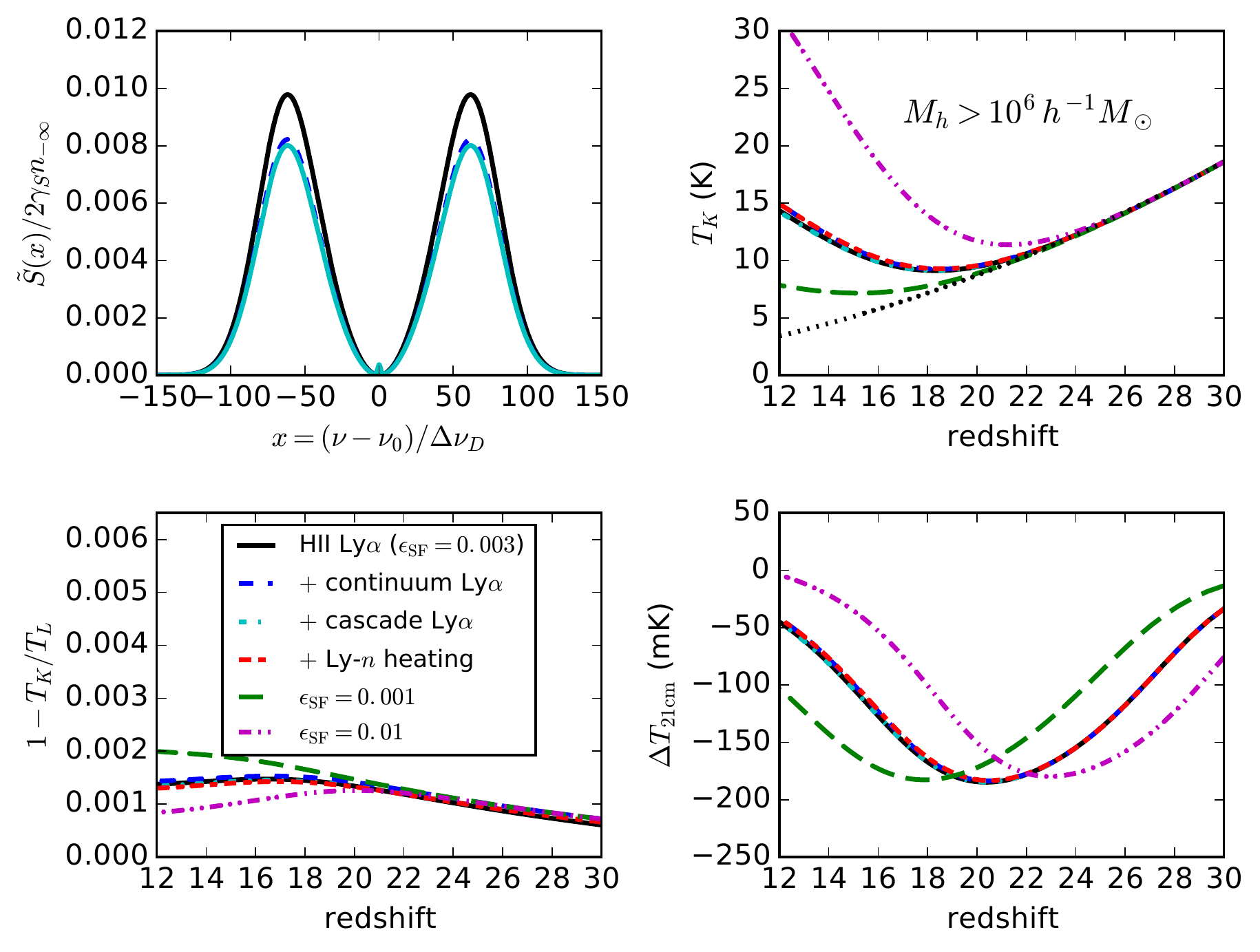}}
\caption{Slab source function at $z=30$ (top left) and evolution of
  the IGM temperature $T_K$ (top right), heating efficiency factor
  $1-T_K/T_L$, where $T_L$ is the light temperature (lower left), and
  the differential 21-cm antenna temperature $\Delta T_{\rm 21 cm}$
  (lower right) for emission from haloes with total masses
  $M_h>10^6\,h^{-1}\,M_\odot$. Internal scattering in the sources
  through a neutral hydrogen column density of
  $10^{18}\,{\rm cm}^{-2}$ with temperature $T_{\rm ISM}=10^4$~K is
  assumed.
}
\label{fig:slab_temperature_evol}
\end{figure}

The impact of the emergent \Lya\ radiation from galaxies on the
differential 21-cm antenna temperature depends both on the intensity
of the radiation and its thermal influence on the IGM through atomic
recoils. The scattering of \Lya\ photons through a static ISM splits
the \Lya\ line into a characteristic double-lobed profile, with horns
symmetrically placed to the red and blue of line centre. To estimate
the amount of heating by these \Lya\ photons, the \Lya\ emission line
is approximated using the emergent flux from a static slab,
\begin{equation}
\tilde S(x_s) = \frac{1}{6^{1/2}}\pi\gamma_S n_{-\infty}\frac{x_s^2}{a\tau_0}\frac{1}{{\rm
    cosh}\left[\exp\left((\pi^2/ 6)(2/3)^{1/2}x_s^3/ (a\tau_0)\right)\right]},
\label{eq:Sxslab}
\end{equation} 
\citep{1973MNRAS.162...43H}, where
$x_s=(\nu-\nu_\alpha)/\Delta\nu_D(T_s)$ for a slab of temperature
$T_s$, as shown in Fig.~\ref{fig:slab_temperature_evol} (top left
panel). The source function has been normalized according to
\begin{equation}
\int_{-\infty}^{x_s}\,dx^\prime_s\,\tilde S(x^\prime_s)  
=\frac{2}{\pi}\gamma_Sn_{-\infty}{\rm atan}\left[\exp\left(\frac{\pi^2}
    {6}\left(\frac{2}{3}\right)^{1/2}\frac{x_s^3}{a\tau_0}\right)\right],
\label{eq:Sxintslab}
\end{equation} 
which satisfies the integral constraint Eq~(\ref{eq:Sxconstraint}) in
the absence of a background continuum ($n_\infty=0$). (When required,
the source is appropriately renormalized in the presence of continuum
produced \Lya\ photons to satisfy the integral constraint.)

On solving for the radiative transfer of the source photons using
Eq.~(\ref{eq:nxevFP_ss}), the blue horn is found to produce a (nearly)
flat \Lya\ continuum across line centre
\citep[cf][]{2009MNRAS.393..949H}, which heats the IGM through \Lya\
recoils. The evolution of the IGM temperature and differential 21-cm
antenna temperature for haloes with total masses
$M_h>10^6\,h^{-1}\,M_\odot$ is shown in
Fig.~\ref{fig:slab_temperature_evol}. Emission through a neutral
hydrogen column density of $10^{18}\,{\rm cm}^{-2}$ with temperature
$T_{\rm ISM}=10^4$~K is assumed, characteristic values for a
Str\"omgren region (see the Appendix). The heating by \Lya\ photon
recoils increases the IGM temperature $T_K$ by $z<21$ (top right
panel). For comparison, the IGM temperature with no heating is shown
as the black dotted line, which decreases monotonically with
decreasing redshift, in contrast to the upturn in the models with
heating. Results are shown for \Lya\ emission line photons produced in
the \HII\ regions of stars forming within the haloes with an
equivalent width of 1000~\AA\ for a star formation efficiency
$\epsilon_{\rm SF}=0.003$ (black solid lines).

An allowance for additional \Lya\ photons from redshifted continuum
radiation, normalized to an \HII-region produced \Lya\ equivalent
width of 1000~\AA, slightly enhances the heating (blue dashed
lines). The volume-averaged enhancement in the number of \Lya\ photons
produced through radiative cascades in the IGM is found to be only 14
percent, and results in a slight reduction in the heating rate,
approximately cancelling the heating produced by redshifted continuum
\Lya\ photons (cyan dot-dot-dashed lines). Continuum photons
redshifted into higher order Lyman resonances are found to increase
the heating rate over the \HII\ region produced \Lya\ photons by
$8-10$ percent, approximately cancelling the cooling by the
cascade-produced \Lya\ photons and nearly restoring the
continuum \Lya\ photon heating rate (red short-dashed-long-dashed
lines). In the remainder of this paper, the effects of higher order
Lyman photons is neglected.

Results for models with $\epsilon_{\rm SF}=0.001$ (green long dashed
lines) and 0.01 (magenta dot-dot-dashed lines) are also shown. All the
models produce a distinctive dip in the 21-cm brightness temperature
as a function of redshift (bottom right panel). The \Lya\ recoil
heating efficiency factor $1-T_K/T_L$ is only weakly sensitive to
$\epsilon_{\rm SF}$ (lower left panel).

\begin{figure}
\scalebox{0.43}{\includegraphics{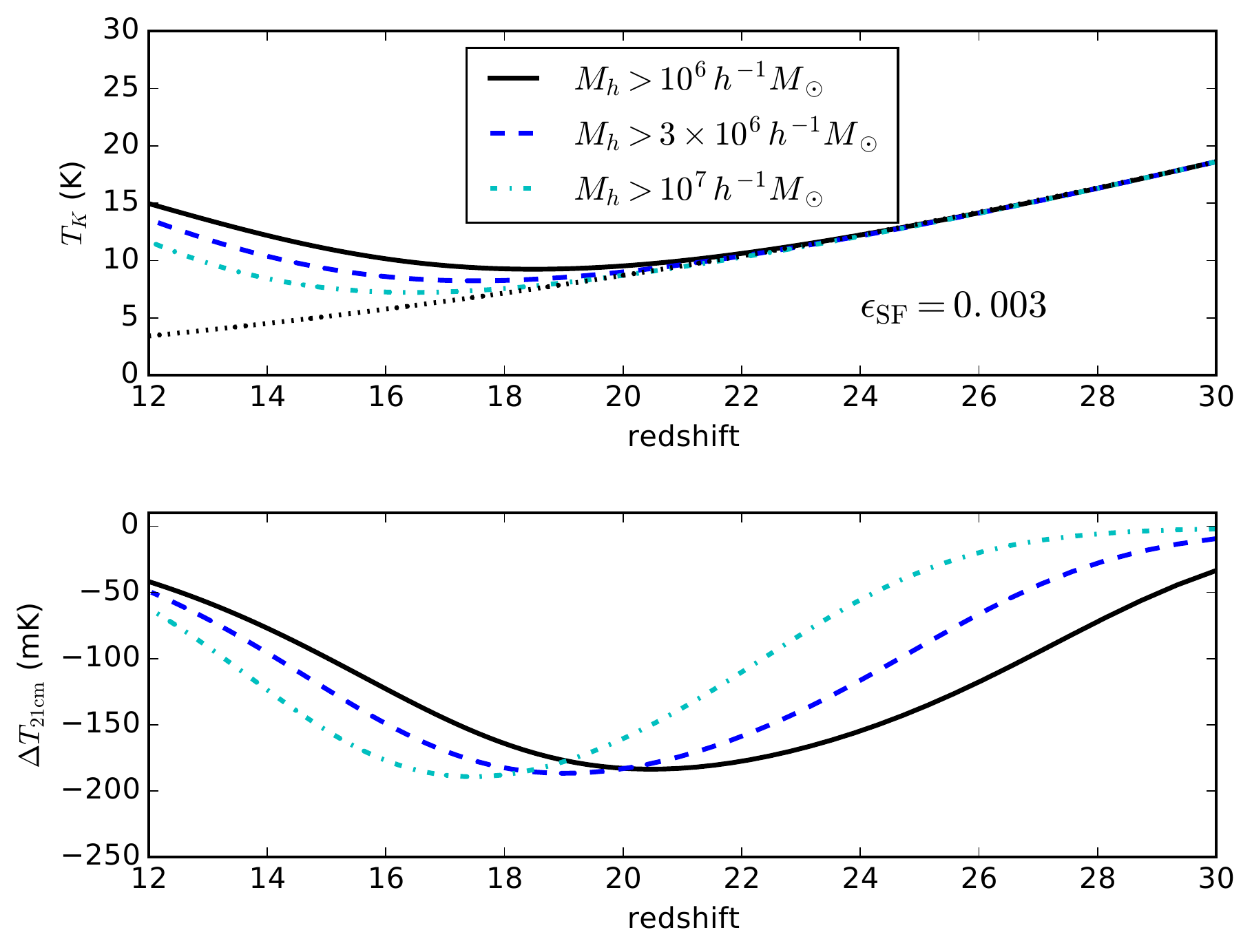}}
\caption{Evolution of the IGM temperature $T_K$ (top panel) and the
  differential 21-cm antenna temperature $\Delta T_{\rm 21 cm}$ (lower
  panel) for minimal halo masses between $M_h=10^6\,h^{-1}\,M_\odot$
  and $10^7\,h^{-1}\,M_\odot$. The star formation efficiency is
  $\epsilon_{\rm SF} = 0.003$. Internal scattering in the sources
  through a neutral hydrogen column density of
  $10^{18}\,{\rm cm}^{-2}$ with temperature $T_{\rm ISM}=10^4$~K is
  assumed.
}
\label{fig:slab_temperature_Mh5t7_evol}
\end{figure}

Allowing for minimal halo masses for star formation between
$10^6-10^7\,h^{-1}\,M_\odot$ shifts the onset of heating, as shown in
Fig.~\ref{fig:slab_temperature_Mh5t7_evol}. A star formation
efficiency of $\epsilon_{\rm SF}=0.003$ is assumed, and the \Lya\
photons scatter through a column density
$N_{\rm HI}=10^{18}\,{\rm cm}^{-2}$ at $T_{\rm ISM}=10^4$~K. The
differential 21-cm antenna temperature reaches a minimum of
$\Delta T_{\rm 21-cm}\simeq-200$~mK, shifting gradually from $z\lta24$
to 17 as the minimum halo mass increases.

\begin{figure}
\scalebox{0.43}{\includegraphics{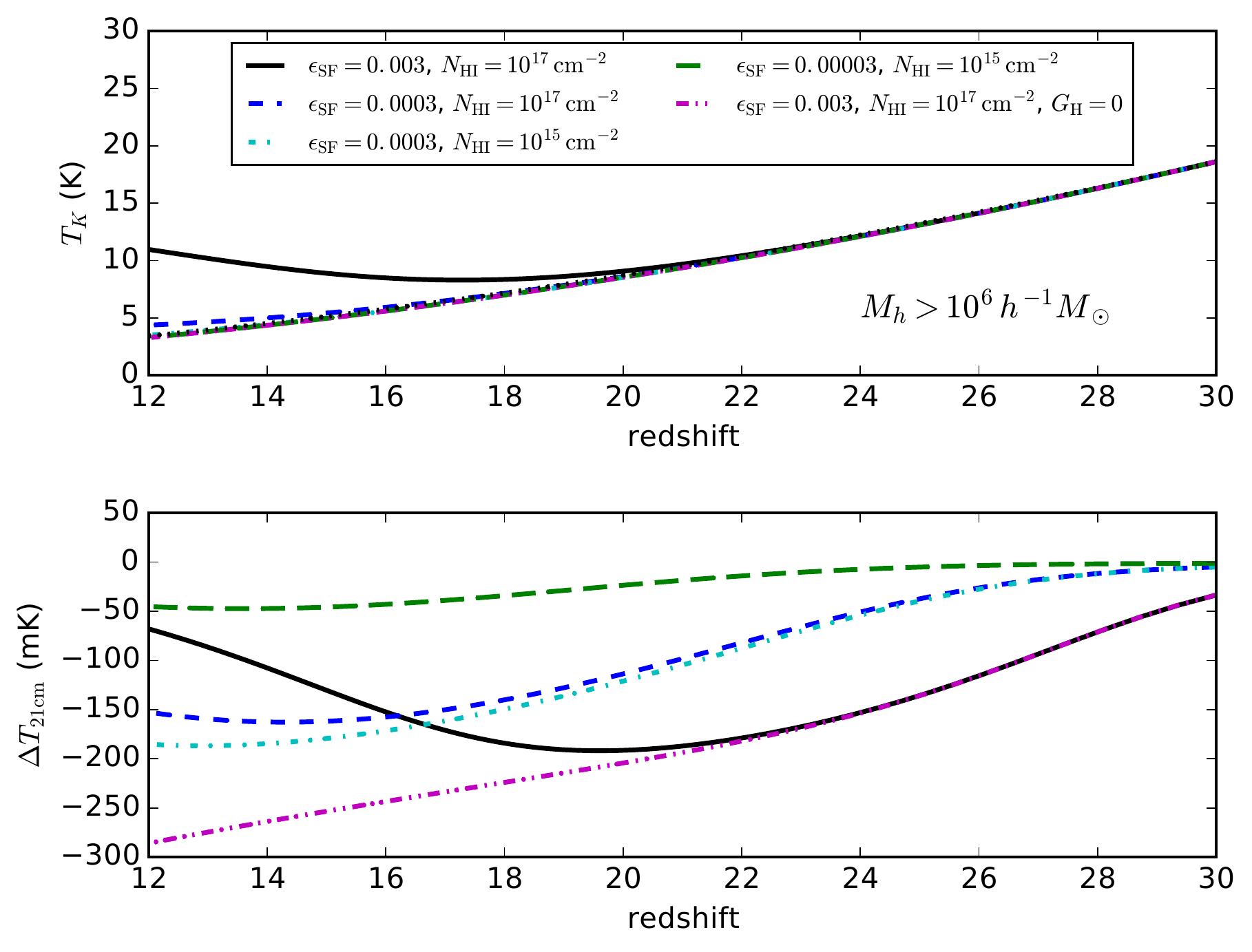}}
\caption{Evolution of the IGM temperature $T_K$ (top panel) and the
  differential 21-cm antenna temperature $\Delta T_{\rm 21 cm}$ (lower
  panel) for a minimal halo mass $M_h>10^6\,h^{-1}\,M_\odot$, star
  formation efficiency $\epsilon_{\rm SF}=0.00003$ to 0.003, and
  internal scattering in the sources through a neutral hydrogen column
  density of $10^{15}-10^{17}\,{\rm cm}^{-2}$. A slab temperature
  $T_{\rm ISM}=10^4$~K is assumed.
}
\label{fig:slab_temperature_NHI1e15t1e17_evol}
\end{figure}

For non-zero escape fractions of ionizing radiation, a Str\"omgren
sphere may not form around the star-forming regions within the
galaxies. In this case, the \HI\ column density through which the
\Lya\ photons scatter need not be as high as $10^{18}\,{\rm
  cm}^{-2}$.
Decreasing the column density to $10^{17}\,{\rm cm}^{-2}$ slightly
lowers the differential antenna temperature, as shown in
Fig.~\ref{fig:slab_temperature_NHI1e15t1e17_evol} (black solid
lines). Since the \Lya\ production rate was not altered, a lower \HI\
column density has the equivalent effect of an enhanced level of \Lya\
production compared with the case of a vanishing escape fraction of
ionizing radiation. This is because the blue lobe of the emergent
\Lya\ spectrum moves closer to line centre and the photons no longer
heat the gas through recoils as efficiently.

By contrast, lowering the star formation rate, and so the \Lya\
production rate, by an order of magnitude reduces the amount of
heating by \Lya\ photon recoils as well as the degree of coupling to
the IGM temperature, resulting in a shallower trough in
$\Delta T_{\rm 21-cm}$ (short-dashed blue lines).

Decreasing the \HI\ column density to $10^{15}\,{\rm cm}^{-2}$ has
little additional effect (cyan dot-dashed lines). It does, however,
introduce net cooling at $z>18.5$, as the redward peak in the \Lya\
emission profile through the slab now moves closer to line
centre. Cooling by recombinant \Lya\ emission line photons produced
within the \HII\ regions now exceeds the heating by
continuum-produced \Lya\ photons. The amount of cooling is slight,
lowering the IGM temperature to $T_K=7.4$~K, compared with the
adiabatic cooling value of 7.6~K. Decreasing the star formation
efficiency by another order of magnitude produces a very shallow dip
in the 21-cm differential antenna temperatures (green long-dashed
lines). For all the models, by $z<18$, the differential antenna
temperature lies well above the minimum level allowing only for
adiabatic expansion and complete coupling of the spin temperature to
the IGM temperature, which approaches $\Delta T_{\rm 21-cm}\gta-300$mK
by $z=12$ (magenta dot-dot-dashed lines).

\subsection{Expanding ISM models}
\label{subsec:expansion}

Observed spectra of \Lya-emitting star-forming galaxies often exhibit
distorted \Lya\ profiles with a dominant redward peak
\citep[eg][]{2010ApJ...717..289S, 2011ApJ...730....5H}, consistent
with an outflow interpretation \citep{2012AA...546A.111V,
  2015MNRAS.449.4336S}. The outflow velocities are typically in excess
of $100\kms$. The velocity offset decreases with inferred halo mass;
extrapolating the trend to low mass haloes corresponds to $4-7\kms$
for halo masses $10^6-10^7h^{-1}M_\odot$ \citep{2018ApJ...856....2M}.

Such low values, however, may not be realizable for \Lya\ photons
originating in \HII\ regions. The Doppler broadening alone is $13\kms$
for gas at $10^4$~K. The flux of \Lya\ photons scattering through a
typical Str\"omgren sphere \HI\ column density of
$10^{18}\,{\rm cm}^{-2}$ peaks at $\sim35\kms$ on the blue and red
sides of line centre \citep{1973MNRAS.162...43H}. An outflow velocity
comparable to the halo escape velocities would little shift the peaks
redward, so that the 21-cm signature would be similar to the static
slab model considered above.

If, however, a residual strong wind subsided to an outflow velocity of
$\sim50\kms$, then the blue-most peak would be shifted to $\sim15\kms$
to the red of line centre. We approximate the resulting \Lya\ photon
emission profile using the homogeneous outflow solution in the
diffusion approximation of \citet{1999ApJ...524..527L}. This model
assumes a point source with a $\delta-$function frequency profile
centred in the red Lorentz wing of \Lya. The \Lya\ photons scatter in
an infinite homogeneous ISM expanding isotropically about the
source. We approximate the emergent \Lya\ emission from a halo using
the flux at a given radius $R_w$ representing the edge of the halo,
with a linear velocity profile out to the edge $v=v_w(r/ R_w)$. The
source function for a source number density $n_s$ is then
\begin{equation}
  S_\nu =n_s(4\pi R_w)^2H_\nu
\label{eq:windsource}
\end{equation}
where the specific flux $H_\nu$ is related to the angle-averaged
intensity $J_\nu$ and ISM inverse attenuation length $\chi_\nu$ by
$H_\nu=-(\partial J_\nu/\partial r)/ (3\chi_\nu)$. It is convenient to
express the solution using the dimensionless frequency and distance
variables $\tilde\nu=(\nu_\alpha-\nu)/\nu_*$ and $\tilde r=r/r_*$,
respectively, where
$\nu_*=\sigma_\alpha A_\alpha\lambda_\alpha n_{\rm HI}/(4\pi^2 dv/dr)$
and $r_*=\lambda_\alpha\nu_*/(dv/dr)$. At $r>r_*$, the diffusion
approximation begins to break down and the \Lya\ photons start to free
stream \citep{1999ApJ...524..527L, 2012MNRAS.426.2380H}. The
dimensionless flux is then
\begin{equation}
\tilde H_{\tilde\nu}=\frac{3\tilde
  r}{8\pi\tilde\nu}\left(\frac{9}{4\pi\tilde\nu^3}\right)^{3/2}e^{-9{\tilde
    r}^2/4{\tilde\nu}^3}.
\label{eq:windflux}
\end{equation}
We normalize the comoving source by
\begin{equation}
\int_{-\infty}^{x_w}\,dx^\prime_w\,\tilde S(x^\prime_w) 
=\frac{2}{\pi^{1/2}}\gamma_Sn_{-\infty}\gamma\left(\frac{3}{2},t\right),
\label{eq:Sxintwind}
\end{equation}
where $x_w=(\nu-\nu_\alpha)/\Delta\nu_D(T_w)$ for an ISM temperature
$T_w$ in the expanding wind,
$t=-(\nu_*/\Delta\nu_D)^39{\tilde R_w}^2/4x^3$, $\gamma(a,t)$ is an
incomplete gamma function and $x<0$. The peak of the emergent flux is
at
\begin{equation}
-x_{\rm
  peak}=\frac{3}{22^{1/3}}\left(\frac{R_w}{r_*}\right)^{2/3}\frac{\nu_*}{\Delta\nu_D}.
\label{eq:windpeak}
\end{equation}
Several criteria must be satisfied for the validity of the expanding
wind solution:\ (1)\ the emission peak must lie in the Lorentz wing,
(2)\ the mean free path at the emission peak must be smaller than
$R_w$, and (3) the wind region must satisfy $R_w<r_*$. 

\begin{figure}
\scalebox{0.43}{\includegraphics{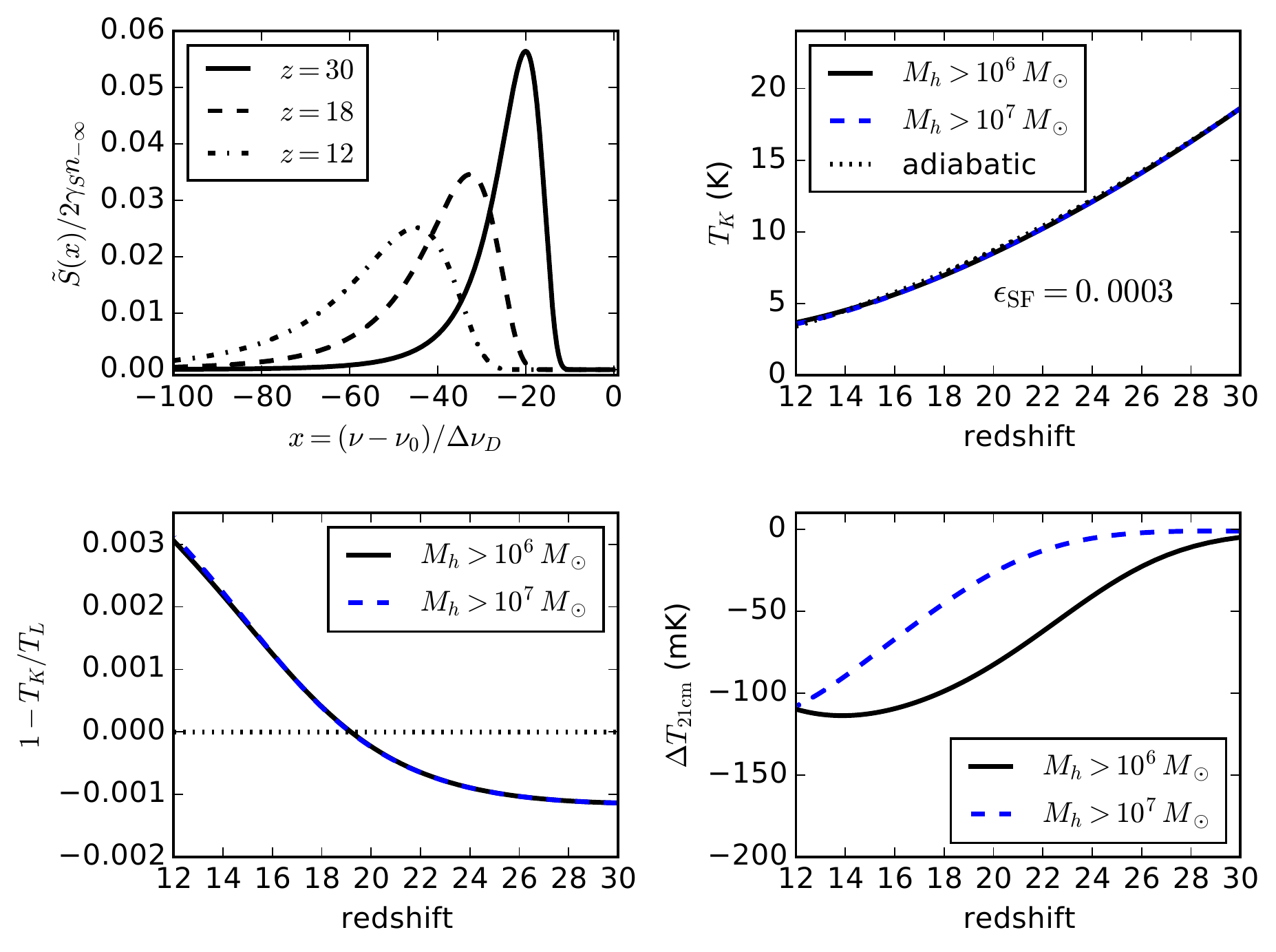}}
\caption{Outflow source function (top left) and resulting evolution of
  the IGM temperature $T_K$ (top right), heating efficiency factor
  $1-T_K/T_L$, where $T_L$ is the light temperature (lower left), and
  the differential 21-cm antenna temperature $\Delta T_{\rm 21-cm}$
  (lower right) for emission from haloes with total masses $M_h>10^6$
  and $10^7\,h^{-1}\,M_\odot$. Internal scattering in the sources
  through a neutral hydrogen density of $10^{-4}\,{\rm cm}^{-3}$ with
  temperature $T_{\rm ISM}=110$~K, an outflow region extending to
  $R_w=200$~pc and maximum outflow velocity $v_w=7\kms$ are assumed.
}
\label{fig:outflow_temperature_evol}
\end{figure}
 
Results from numerical simulations show the gaseous contents of the
haloes of primordial galaxies to be highly dynamic. The first
supernovae in a halo will expel the halo gas in a wind. The hydrogen
density may be lowered to values below $10^{-4}\,{\rm cm}^{-3}$ with
temperatures cooling to $100-1000$~K \citep{2008ApJ...682...49W,
  2013ApJ...774...64W, 2015MNRAS.452.2822S}. Subsequently, gas that
has not escaped may refill the haloes, resulting in a back flow and
re-establishing hydrogen densities of $\sim1\,{\rm cm}^{-3}$. Inflow
from mergers will also contribute to the restoration of the gaseous
contents. These effects are still poorly understood, with no direct
observations.

For a neutral hydrogen density $1\,{\rm cm}^{-3}$ and temperature
$10^4$~K in a halo with a linear outflow velocity profile reaching
$50\kms$ at $R_w=100$~pc, $x_{\rm peak}\simeq-670$ for an IGM
temperature of 18~K. Such a redward peak will have little effect on
the IGM; the interaction between the sources and the IGM will be
dominated by the source continua. Lowering the \HI\ density to as low
as $0.004\,{\rm cm}^{-3}$ (holding the halo gas temperature and flow
parameters fixed), still results in a peak at
$x_{\rm peak}\simeq-110$. Lower densities violate the conditions
required for the diffusion approximation.

In the absence of a high \HI\ column density, a large outflow velocity
is no longer required for validity of the expansion solution. For a
200~pc radius with an expansion velocity of $7\kms$, on the order of a
halo escape velocity, a hydrogen density of $10^{-4}\,{\rm cm}^{-3}$
and ISM temperature of 110~K results in $x_{\rm peak}\simeq16-40$ over
$z=30$ to 12. (Too high an ISM temperature, exceeding 600~K, results
in a peak within the Doppler core.) The absence of a high \HI\ column
density, however, requires a low \Lya\ photon production rate to
ensure the escaping Lyman continuum (LyC) radiation does not overly
ionize the IGM (see Fig.~\ref{fig:Pana_evol}).

Results are shown in Fig.~\ref{fig:outflow_temperature_evol} for
$\epsilon_{\rm SF}=0.0003$ and a \Lya\ equivalent width of
$W=1000$~\AA\ relative to the continuum. The strong red peak in the
source function initiates cooling of the IGM by \Lya\ scattering
(lower left panel), with the IGM temperature decreasing slightly
faster than given by adiabatic expansion alone, but at most by about 5
percent (top right panel). Because the IGM temperature decreases with
time (from cooling both by \Lya\ recoils and adiabatic expansion), the
peak in the source function shifts increasingly towards the red,
reducing the strength of the cooling. By $z\simeq19$, recoil heating
by the \Lya\ continuum dominates. Despite the lower IGM temperatures,
the differential 21-cm absorption signature (lower right) is weaker
than for the static slab models considered above because the \Lya\
scattering rate is smaller.

\subsection{Super-adiabatic cooling models}
\label{subsec:superadiabatic}

The \Lya\ scattering induced cooling in
Fig.~\ref{fig:outflow_temperature_evol} illustrates that the reddened
\Lya\ radiation spectrum in an expanding ISM is able to cool the IGM
faster than adiabatic expansion alone. The cooling rate is limited by
the requirement that the escaping LyC radiation from the source
population does not overly photoionize the IGM; it is not limited by
the \Lya\ photon flux. Motivated by the low differential 21-cm antenna
temperatures reported from the EDGES experiment
\citep{2018Natur.555...67B}, we explore how large a cooling rate by
\Lya\ photon recoil scattering may be achieved, presuming a low
intrinsic level of LyC ionizing radiation.

\begin{figure*}
\scalebox{0.86}{\includegraphics{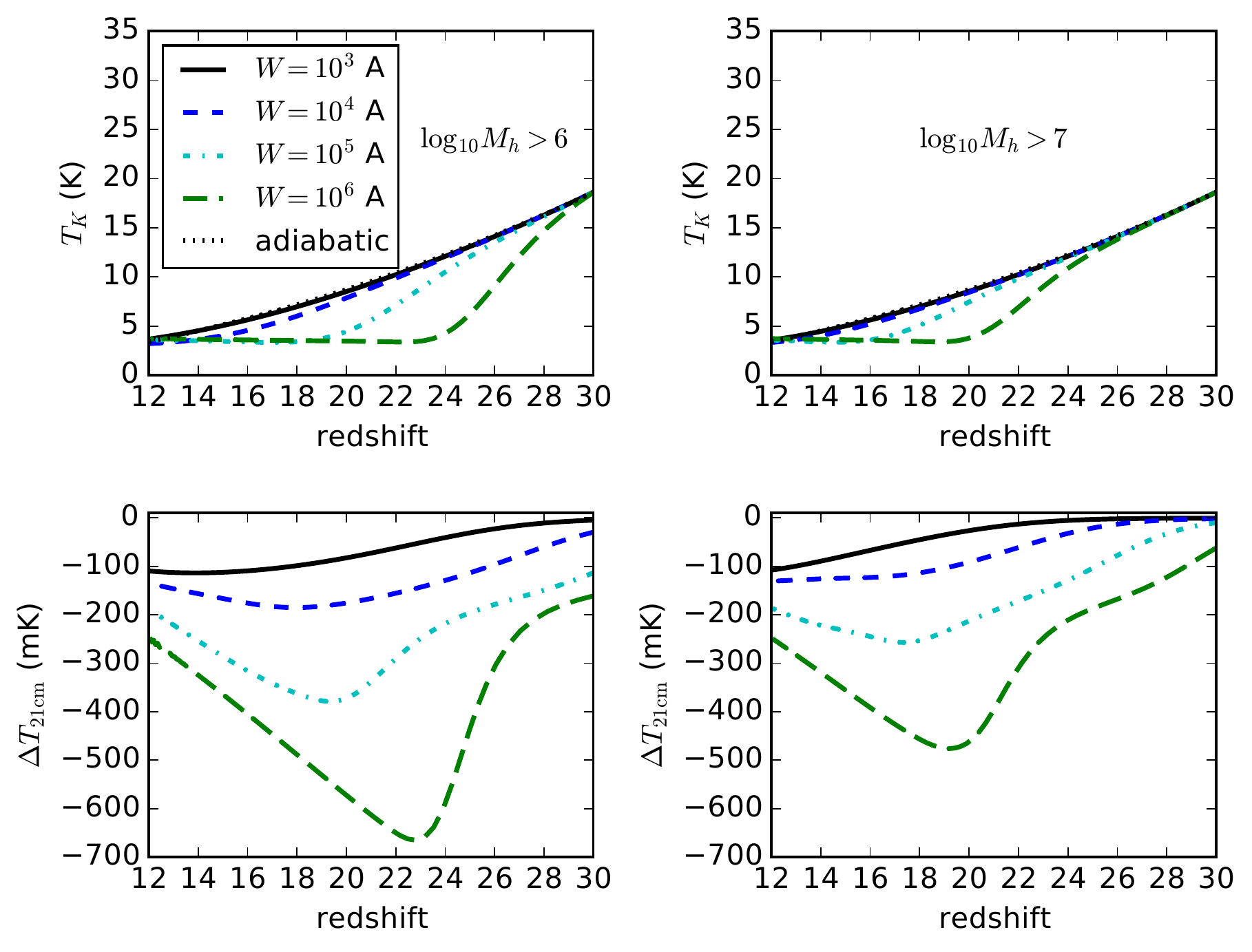}}
\caption{Evolution of the IGM temperature $T_K$ (top panels), and the
  differential 21-cm antenna temperature $\Delta T_{\rm 21-cm}$
  (bottom panels) for emission from haloes with total masses
  $M_h>10^6$ (left hand panels) and $10^7\,h^{-1}\,M_\odot$ (right
  hand panels), with star formation efficiency factor
  $\epsilon_{\rm SF}=0.0003$. Internal scattering through an
  outflowing region extending to $R_w=200$~pc and maximum outflow
  velocity $v_w=7\kms$ with a neutral hydrogen density of
  $10^{-4}\,{\rm cm}^{-3}$ and temperature $T_{\rm ISM}=110$~K is
  assumed. A central \Lya\ source is assumed present with equivalent
  widths of $W=10^3$~\AA, $10^4$~\AA, $10^5$~\AA\ and $10^6$~\AA,
  considered, as indicated.
}
\label{fig:outflow_temperature_evol_ew1e3t6}
\end{figure*}
 
In Fig.~\ref{fig:outflow_temperature_evol_ew1e3t6}, the IGM
temperature and differential 21-cm antenna temperature evolution are
shown for a star formation efficiency $\epsilon_{\rm SF}=0.0003$, for
which the LyC radiation from haloes with masses exceeding
$10^6\,h^{-1}M_\odot$ would not be sufficient to photoionize the IGM
at a level in conflict with CMB limits on the intergalactic Thomson
optical depth. The same cool, low density slowly expanding internal
environment is assumed for the haloes as in
Sec.~\ref{subsec:expansion}.

Results are shown for a central source producing \Lya\ emission line
photons with equivalent width values in the range $W=10^3-10^6$~\AA. A
value of 4000~\AA\ is at the upper end of plausible \Lya\ equivalent
widths produced by \HII\ regions ionized by Pop~III stars
\citep{2010AA...523A..64R}, so the higher values considered greatly
exceed the expectation from radiative recombinations in photoionized
gas. There are no astrophysical models for the origin of such high
\Lya\ equivalent widths; it is noted that other physical mechanisms
are available for producing \Lya\ photons, including the generation of
\Lya\ photons through electron collisional excitation of neutral
hydrogen in shock fronts propagating through largely neutral gas and
through the impact of soft X-rays in nearly neutral gas followed by
excitations by secondary electrons. With a normalizing \Lya\
equivalent width of 1000~\AA\ for $\epsilon_{\rm SF}=0.0003$,
equivalent widths of $10^4-10^6$~\AA\ correspond to energy conversion
efficiencies of the halo baryons into \Lya\ photons of approximately
$10^{-5}-10^{-3}$. As an example, for accretion onto a black hole the
upper end corresponds to a typical black hole energy conversion
efficiency of $0.1-0.3$.

As shown in the top right hand panel of
Fig.~\ref{fig:outflow_temperature_evol_ew1e3t6}, intense \Lya\
emission is able to cool the IGM substantially faster than adiabatic
cooling from the expansion of the Universe. Much lower differential
21-cm antenna temperatures are produced, extending to as low as
$\Delta T_{\rm 21-cm}\sim-700$mK, with values
$\Delta T_{\rm 21-cm}<-300$~mK possible over the redshift range
$13<z<26$. The signal diminishes towards lower redshift because the
\Lya\ photons eventually heat the gas. In fact a floor IGM temperature
of $T_K=3.2$~K is found
(Fig.~\ref{fig:outflow_temperature_evol_ew1e3t6}, upper panels), with
\Lya\ cooling converting to heating at $T_K\lta4$~K.

The expanding wind model applies only for radiation in the Lorentz
wing of the \Lya\ scattering cross section. Cooling may also occur for
scattering across the Doppler core for a sufficiently red source. This
is demonstrated by modelling a reddened emission spectrum from a slab
as a double-Gaussian profile of the form
\begin{equation}
  \tilde S(x) = a_0(1)e^{-[x-x_0(1)]^2/2\sigma^2_0(1)} +
  a_0(2)e^{-[x-x_0(2)]^2/2\sigma^2_0(2)},
\label{eq:Sx_DG}
\end{equation}
normalized according to
\begin{eqnarray}
\frac{4}{\gamma_Sn_{-\infty}}\int_{-\infty}^x\,dx^\prime \tilde
  S(x^\prime)&=&\left[1+{\rm erf}\left(\frac{x-x_0(1)}{2^{1/2}\sigma_0(1)}\right)\right]\nonumber\\
&+&\left[1+{\rm erf}\left(\frac{x-x_0(2)}{2^{1/2}\sigma_0(2)}\right)\right],
\label{eq:Sxint_DG}
\end{eqnarray}
in agreement with Eq~(\ref{eq:Sxconstraint}) in the absence of a
background continuum ($n_\infty=0$). The line widths correspond to the
temperature of the emitting gas. Reddening through a surrounding
expanding circum-galactic medium is modelled by assuming an asymmetric
profile.

\begin{figure}
\scalebox{0.43}{\includegraphics{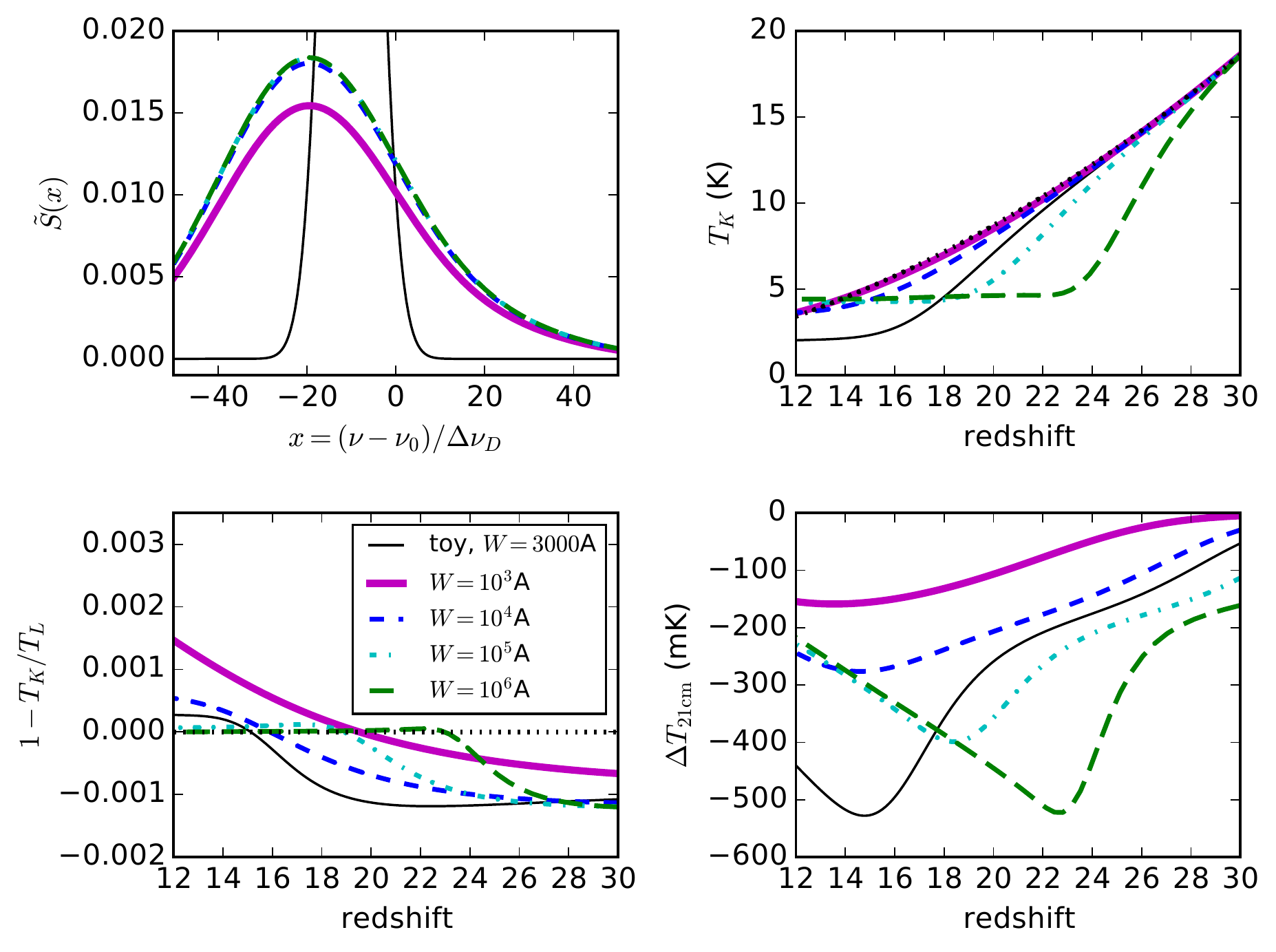}}
\caption{Evolution of IGM temperature including \Lya\ photon recoil
  cooling and heating for double-Gaussian source profiles, with
  initial centres at $\pm20\Delta\nu_{D, i}$ and initial widths of
  $20\Delta\nu_{D,i}$, where $\Delta\nu_{D, i}$ is the Doppler width
  of the IGM at $z_i=30$. The red-to-blue amplitudes of the Gaussians
  are in the ratio 10:1. Emission is from haloes with
  $M_h>10^6\,h^{-1}M_\odot$ and $\epsilon_{\rm SF}=0.0003$, and \Lya\
  equivalent widths from $W=10^3-10^6$~\AA. Shown are:\ (upper left)
  normalized source profiles, (upper right) evolution in IGM kinetic
  temperature, (lower left) \Lya\ recoil heating efficiency
  $1-T_K/T_L$, (lower right) differential 21-cm antenna
  temperature. Also shown for comparison are results for a toy model
  with a sharp \Lya\ Gaussian source centred at $x=-10$, normalized to
  $W=3000$~\AA\ (thin black lines).
}
\label{fig:double-gaussian}
\end{figure}

Allowing for the freedom to place the peak of a red profile closer to
line-centre, we first examine how much cooling may be initiated by a
moderate equivalent width emission line. We consider a toy model:\ a
sharp pure red Gaussian profile initially centred at $x_0=-10$ with an
initial width $\sigma_0=5$ at $z=30$. The source is normalized to an
equivalent width $W=3000$~\AA\ for a star formation efficiency
$\epsilon_{\rm SF}=0.002$, sufficiently low to ensure the associated
LyC radiation will not overly photoionize the IGM, assuming a minimum
halo mass $M_h>10^6\,h^{-1}M_\odot$ for the sources. As shown in
Fig.~\ref{fig:double-gaussian} (thin black lines), strong cooling is
achieved, with a minimum 21-cm brightness temperature of
$\Delta T_{\rm 21-cm}\sim-500$~mK reached by $z=15$. The position of
the source peak corresponds to an expansion velocity of $\sim4\kms$
and the width to thermal broadening of an ISM at 460~K. Such a \Lya\
source could not be directly generated by radiative recombinations or
collisional excitation, although possibly such photons may be
generated in cold gas by secondary electrons from X-rays impinging on
such cold gas, if such a scenario is workable.

A case with a more plausible ISM temperature is illustrated with an
initial $[x_0(1), x_0(2)]=[-20,20]$ and $\sigma_0(1)=\sigma_0(2)=20$
at $z=30$, corresponding to thermally broadened Doppler emission
profiles from a medium with temperature
$20^2T_{\rm IGM}(z_i=30) = 7400$~K. The Gaussian centres and widths
are fixed in physical units. A red-to-blue amplitude ratio 10:1 is
assumed. The resulting evolution in gas temperature and differential
21-cm antenna temperature is shown in Fig.~\ref{fig:double-gaussian}
for emission from haloes with $M_h>10^6\,h^{-1}M_\odot$,
$\epsilon_{\rm SF}=0.0003$ and \Lya\ equivalent widths of
$10^3-10^6$~\AA. Similar evolution in the IGM temperature is found as
for the expanding sphere model, with a floor temperature of $\sim4$~K
before \Lya\ recoil cooling converts to heating. A minimum
differential 21-cm antenna temperature as low as $\sim-500$~mK may be
reached.

\section{Discussion}
\label{sec:discussion}

Full coupling of the 21-cm spin temperature to the IGM temperature,
allowing for adiabatic expansion alone of the IGM since the
recombination era, produces a differential 21-cm antenna temperature
$\Delta T_{\rm 21-cm}$ compared with the CMB ranging from -170~mK at
$z=30$ to -290~mK at $z=12$. For conventional star formation rate
assumptions in the first galaxies, the \Lya\ intensity is not expected
to be sufficiently strong to produce strong coupling between the IGM
spin and kinetic temperatures until $z\lta20$, with
$\Delta T_{\rm 21-cm}\simeq-10$~mK at $z=30$, and decreasing to
$\Delta T_{\rm 21-cm}<-200$~mK only by $z<20$
\citep[eg][]{2005ApJ...626....1B, 2006MNRAS.367..259H}.

We consider a broader class of models, allowing for star formation in
haloes with minimum masses of $10^6-10^7\,h^{-1}M_\odot$, and explore
a range of assumptions concerning the conversion efficiency of baryons
to stars, and the internal gas structure of the haloes. We also
consider the enhanced energy production of LyC and \Lya\ photons in
\HII\ regions within primordial galaxies dominated by Pop~III stars,
following \citet{2010AA...523A..64R}. \Lya\ emission equivalent widths
of $1000-4000$~\AA\ may result, several times larger than predictions
for the \HII\ regions of Pop~II stars.

For high escape fractions of LyC photons into the IGM, many of the
models predict an amount of ionization of the IGM inconsistent with
CMB limits on the intergalactic Thomson optical depth. The high escape
fractions also result in reduced \Lya\ photon production within the
\HII\ regions of the galaxies. For low escape fractions instead, an
intense rate of \Lya\ production results, sufficient to alter the
intergalactic temperature through recoils between \Lya\ photons and
the neutral intergalactic hydrogen.

In the simplest model of a small LyC radiation escape fraction through
a static ISM, the \HI\ column density within the \HII\ regions is at
least $\sim10^{18}\,{\rm cm}^{-2}$. Approximating the radiative
transfer through this column density by a static slab model produces
two broadly spaced \Lya\ emission horns emergent from the
galaxies. Solving for the intergalactic radiative transfer of the
\Lya\ photons from these galactic sources shows that, for a
conservative estimate of the conversion efficiency of baryons to
stars, the blue horns produce a strong metagalactic \Lya\ continuum
across line-centre that acts as a heat source through \Lya\ photon
recoils and slows the adiabatic cooling of the IGM. The \Lya\
metagalactic field more strongly couples the spin temperature to the
IGM kinetic temperature than in conventional models. A floor
differential antenna temperature of
$\Delta T_{\rm 21-cm}\simeq-200$~mK results for this class of
models. This is because the heating of the IGM by \Lya\ recoils
eventually causes the 21-cm signal to diminish, producing a
characteristic trough in $\Delta T_{\rm 21-cm}$ between $12<z<30$. For
a minimal halo mass for forming stars of $10^6\,h^{-1}M_\odot$,
$\Delta T_{\rm 21-cm}$ reaches $\sim-200$~mK by $z=21$.  For a
minimal halo mass of $10^7\,h^{-1}M_\odot$, the minimum is not reached
until $z\sim18$, as in conventional models. The redshift for reaching
$\Delta T_{\rm 21-cm}\simeq-200$~mK is shifted to earlier epochs for
more efficient baryon conversion into stars.

Decreasing the star formation efficiency by an order of magnitude
avoids over-ionizing the IGM without requiring an optically thick \HI\
column density (at the Lyman edge). For a low density ISM, the \HII\
regions can continue to grow without becoming recombination bound. In
models in which the interior halo gas expands at velocities comparable
to the escape velocity, as may result following a massive supernova
explosion, \Lya\ photons produced by a subsequent episode of star
formation would be reddened as they scatter through the expanding halo
gas. In such a scenario, much more moderate 21-cm signatures result,
comparable to more conventional scenarios for a minimal halo mass for
star formation of $10^7\,h^{-1}M_\odot$. For a minimal halo mass as
low as $10^6\,h^{-1}M_\odot$, $\Delta T_{\rm 21-cm}$ reaches a minimum
of $\sim-120$~mK by $z=14$. At lower redshifts, the metagalactic \Lya\
field is sufficiently strong to heat the IGM, reducing the 21-cm
signature to $\Delta T_{\rm 21-cm}\simeq-100$~mK by $z=12$.

At $z>19$, the models with an expanding interior actually cool the IGM
through \Lya\ photon recoils. Enhancing the \Lya\ emission equivalent
width enhances the cooling. We explore the consequences of models with
\Lya\ emission equivalent widths reaching $10^6$~\AA. Such extreme
values are not predicted from radiative recombinations within any
existing models; but neither are such values astrophysically ruled
out. Such intensities may be produced through an alternative route,
such as collisional excitation within shock fronts or by secondary
electrons in neutral gas produced by soft X-rays impinging on the
surroundings of the sources. Whether it is possible to achieve such
high equivalent widths whilst still producing a predominantly reddened
\Lya\ spectrum or without excessive heating by X-rays in the large
scale IGM is unknown.

We show that intense \Lya\ production with such extreme equivalent
widths is able to supercool the IGM, lowering the gas kinetic
temperature to a floor of $3-4$~K before the \Lya\ photons begin to
heat the gas through recoils. A broad dip in the 21-cm antenna
temperature results over redshifts $12<z<30$, with a minimum
temperature as low as $\Delta T_{\rm 21-cm}\sim-700$~mK. The signature
is similar to the measurement reported for the EDGES experiment
\citep{2018Natur.555...67B}, including a sharp decline as the
signature develops, the sharpness increasing as the threshold halo
mass for forming stars is increased. This is because the number of
haloes grows more rapidly with time the greater the threshold halo
mass. Although the signal decreases much more gradually than does the
EDGES signal, the signal would decrease more rapidly if additional IGM
heating sources formed, such as X-ray emitting compact binaries within
the primordial galaxies.

\section{Conclusions}
\label{sec:conclusions}

Within the context of conventional models of star formation in
primordial galaxies, the differential 21-cm antenna temperature
between the IGM and the CMB is expected to lie between
$-100<\Delta T_{\rm 21-cm}<0$ at $z>20$. In this paper, we extend the
models to explore a broader range of possible 21-cm signatures at
these early epochs.

Allowing for enhanced \Lya\ photon production from Pop~III star \HII\
regions \citep{2010AA...523A..64R}, we find the \Lya\ scattering rate
$P_\alpha$ may exceed the thermalization rate $P_{\rm th}$ required to
begin coupling the 21-cm spin temperature to the gas kinetic
temperature as early as redshifts $30-25$ for a minimum halo mass for
star formation between $10^6-10^7\,h^{-1}M_\odot$. Such high levels of
\Lya\ emission, however, may be accompanied by a metagalactic LyC
background sufficiently strong to photoionize the IGM at a level in
excess of the CMB limits on the Thomson optical depth.

The excessive reionization may be suppressed in two ways, by
exhausting the supply of ionizing photons within recombination-bound
\HII\ regions of the primordial galaxies or by reducing the star
formation efficiency. We examine both cases. For the latter, we also
allow for expanding interior gas within the haloes that may redden the
emergent \Lya\ spectrum, motivated by observations of reddened \Lya\
spectra \citep[eg][]{2010ApJ...717..289S, 2011ApJ...730....5H}, and by
models of the impact of supernovae on the gaseous interiors of
primordial haloes \citep{2008ApJ...682...49W, 2013ApJ...774...64W,
  2015MNRAS.452.2822S}.

The \Lya\ photons produced within a recombination-bound \HII\ region
must traverse an \HI\ column density of $\sim10^{18}\,{\rm cm}^{-2}$
or more. A characteristic double-horned profile results, displaced by
several Doppler widths (referenced to the IGM temperature) to the blue
and red of line-centre. The resulting \Lya\ metagalactic radiation
field acts as an additional heating source. The differential 21-cm
antenna temperature is then limited to a floor of
$\Delta T_{\rm 21-cm}>-200$~mK over $z<30$. The floor arises because
either the \Lya\ radiation field is not sufficiently strong to fully
couple the 21-cm spin temperature to the gas kinetic temperature, or
because the radiation field is so strong it heats the IGM through
\Lya\ photon recoils. The result is an extended trough in the
differential antenna temperature over the redshift range $12<z<30$.

Lowering the star formation efficiency, as may be expected in the
presence of substantial feedback, diminishes the strength of the 21-cm
absorption signature, in agreement with previous estimates. Allowing
for reddening within an expanding halo, however, produces net cooling
by \Lya\ photon recoils, even accounting for the accompanying
continuum-produced \Lya\ photons, which always heat the gas. For
plausible \Lya\ emission line equivalent widths of primordial \HII\
regions, limited to $<4000$~\AA, the amount of cooling is small,
producing less than a 5 percent decrease in the IGM kinetic
temperature compared with the reduction by adiabatic expansion cooling
since the recombination era. By $z<19$, continuum \Lya\ heating
dominates and the IGM begins to warm. The effect is again a dip in the
differential 21-cm antenna temperature, with a minimum
$\Delta T_{\rm 21-cm}>-150$~mK between $12<z<18$, depending on the
minimum halo mass for star formation.

Motivated by the large absorption signature reported from the EDGES
experiment of $\Delta T_{\rm 21-cm}\simeq-650 - -450$~mK between
$16<z<19$ \citep{2018Natur.555...67B}, we also consider models that
may supercool the IGM. Whilst sources emitting to the red of
line-centre with moderate equivalent widths may produce brightness
temperatures as low as $\Delta T_{\rm 21-cm}\sim-500$~mK, they require
narrow widths corresponding to ISM temperatures much smaller than
$10^4$~K; a mechanism for producing \Lya\ photons in such cold gas is
problematic.  Warmer ISM temperatures require instead extreme \Lya\
emission equivalent widths for supercooling, up to $10^6$~\AA, in
expanding halo gas. We find deep absorption signatures of up to
$-700$~mK are achievable for these models, with a broad trough where
$\vert\Delta T_{\rm 21-cm}\vert < 300$~mK over $13<z<26$. As the
redshift decreases, heating by \Lya\ recoils dominates, decreasing the
strength of the absorption signal to
$\vert\Delta T_{\rm 21-cm}\vert< 250$~mK by $z=12$. An astrophysical
model that would produce the required high \Lya\ emission equivalent
widths, however, remains elusive.

\appendix

\section{Constraints on \Lya\ photon scattering rate from reionization
  limits}
\label{app:PLyaReion}

The production rate $Q(H)$ of LyC photons, allowing for an escape
fraction $f_{\rm esc}$, is related to $\dot n_\alpha$ by
\begin{equation}  
Q(H) = \frac{\dot n_\alpha}{(1-f_{\rm esc})P\bar f_{\rm coll}},
\label{eq:QHndota}
\end{equation}  
where $P$ is an energy enhancement factor, increasing with hardness of  
the stellar spectrum and the gas density of the \HII\ region and  
ranging over $1\lta P\lta4$ \citep{2010AA...523A..64R}. The relative  
number of \Lya\ photons produced per recombination is quantified by  
$f_{\rm coll}$, with $2/3 \lta f_{\rm coll}\lta1$. For a small escape  
fraction $f_{\rm esc}<<1$, $2/3 < Q(H)/\dot n_\alpha<4$, so that the  
cumulative numbers of \Lya\ and LyC photons produced are  
comparable. The LyC photons are largely confined to the  
\HII\ regions, producing Str\"omgren spheres, with recombinations  
balancing photoionizations. For an assumed source spectrum varying as  
$\nu^{-\alpha_S}$, the photoionization rate per hydrogen atom a  
distance $r$ from the source is  
$\Gamma_i\simeq [\alpha_S/(3+\alpha_S)]Q(H)\sigma_L/ (4\pi r^2)$, 
where $\sigma_L$ is the photoelectric cross section at the ionization  
threshold energy. For a uniform density gas, the accumulated \HI\ 
column density through the Str\"omgren sphere is then  
\begin{equation} 
N_{\rm HI}^{\rm Str}\approx\frac{3+\alpha_S}{\alpha_S}\frac{1}{\sigma_L}\approx10^{18}\,{\rm cm}^{-2}.  
\label{eq:NHIStrom}
\end{equation} 
For a characteristic temperature $T=10^4T_4$~K for photoionization
heated gas, the line centre optical depth for \Lya\ scattering is $\tau_\alpha\simeq6\times10^4T_4^{-1/2}$. The \Lya\ photons escape the \HII\ region by diffusing in frequency until the \HII\ region is optically thin to \Lya\ photon scattering.  
 
For larger escape fractions, the stars begin to photoionize  
the IGM, with photons entering the IGM at the rate $Q_{\rm  
  IGM}(H)=f_{\rm esc}Q(H)$, so that  
\begin{equation} 
Q_{\rm IGM}(H) = \frac{f_{\rm esc}}{(1-f_{\rm esc}) P\bar f_{\rm coll}}\dot n_\alpha.  
\label{eq:QIGMHndota}
\end{equation} 
The number density of LyC photons is then related to the total number
density $n_\alpha$ of \Lya\ photons produced by
$n_L = f_{\rm esc}n_\alpha /[(1-f_{\rm esc})P\bar f_{\rm coll}]$. For
$f_{\rm esc}=0.2$, for example, if $n_\alpha>\bar n_{\rm H}$, where
$\bar n_{\rm H}$ is the mean IGM hydrogen density, a substantial
fraction of the IGM will have been
photoionized.\footnote{Recombinations within a clumped IGM will slow
  the reionization process, but this is expected to produce only a
  small to moderate delay \citep[eg][]{2012ApJ...747..100S,
    2020arXiv200202467D}.} This may conflict with CMB limits on the
Thomson optical depth. For a 1$\sigma$ (2$\sigma$) upper limit of
$\tau_e<0.06$ (0.07) \citep{2018arXiv180706209P}, the additional
contribution to the optical depth must be limited to
$\Delta\tau_e<0.02$ (0.03) after subtracting the contribution between
$z=0$ and 6. As shown in Fig.~\ref{fig:Pana_evol} (for
$P\bar f_{\rm coll}=2$), this limit may be reached for a minimum halo
mass for star formation of $M_h>10^6\,h^{-1}\,{\rm M}_\odot$ at
$z<13$, assuming the number density of ionized hydrogen and singly
ionized helium atoms matches $n_L$. Since $\tau_e \bar n_H/n_L$
generally exceeds 0.01 at $z<35$, (bottom right hand panel of
Fig.~\ref{fig:Pana_evol}), once $n_L>\bar n_{\rm H}$, the Thomson
optical depth will have reached 0.01. The IGM may thus become close to
being fully photoionized before exceeding the CMB limit. The emission
of LyC radiation from galaxies, however, will have to abruptly cease
to allow the ionized gas to recombine to avoid exceeding the CMB
limit. At $z=15$, the recombination time of $10^4$~K is too long, of
order 100~Myr, corresponding to an additional increment
$\Delta\tau_e\approx0.04$. The Compton cooling time, however, is very
short, a little under 35~Myr. Upon cooling to the CMB temperature, the
gas will quickly recombine. The incremental increase in the optical
depth is then only $\Delta\tau_e\approx0.017$ if fully ionized. It
would thus be possible for about half of the IGM to become ionized
briefly by $z=15$ without exceeding the CMB limit on the Thomson
optical depth. More generally, however, at higher redshifts
$n_L/\bar n_{\rm H}\lta0.5$ is required to ensure the CMB limit is not
exceeded.
 
\begin{figure}
\scalebox{0.47}{\includegraphics{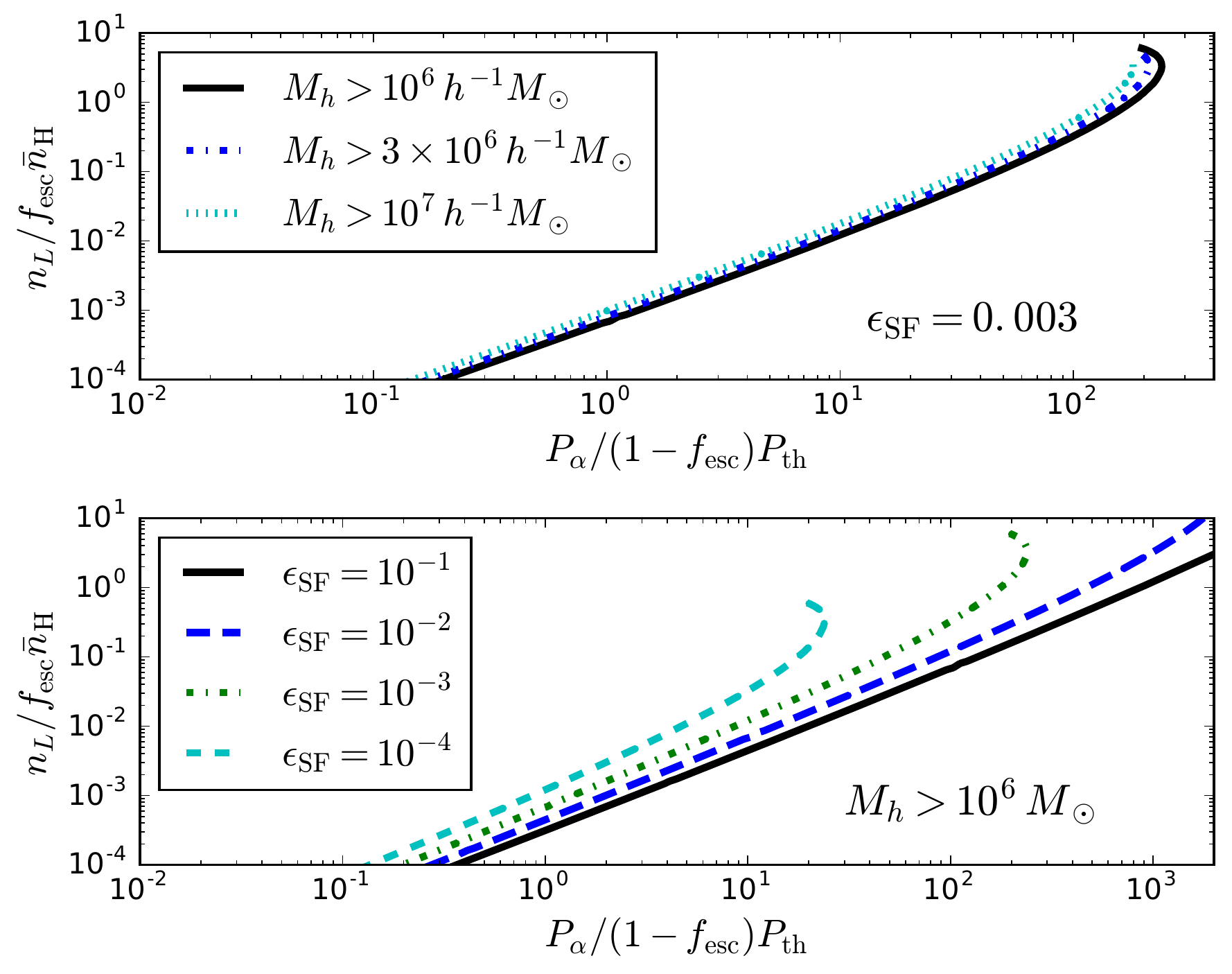}}
\caption{The integrated number density $n_L$ of LyC photons produced
  in the IGM for $f_{\rm esc}>0$, normalized by the total hydrogen
  density, against the \Lya\ scattering rate $P_\alpha$, normalized by
  the thermalization rate $P_{\rm th}$. The top panel shows the
  variation for different minimum halo mass threshold values for star
  formation, and the bottom panel shows the dependence on nuclear star
  formation efficiency $\epsilon_{\rm SF}$ for a halo mass threshold
  $M_h>10^6\,h^{-1}\,{\rm M}_\odot$.
}
\label{fig:PaVna}
\end{figure}
 
From Fig.~\ref{fig:PaVna} (upper panel), for an escape fraction
$f_{\rm esc}<0.2$, requiring $n_L/\bar n_{\rm H}<0.5$ imposes the
restriction $P_\alpha/P_{\rm th}\lta200$ for
$\epsilon_{\rm SF}=0.003$. Even for the extreme efficiency
$\epsilon_{\rm SF}=0.1$, the cumulative number of ionization proceeds
much more quickly and Fig.~\ref{fig:PaVna} (lower panel) shows
$P_\alpha/P_{\rm th}<3000$ (at an earlier epoch) is required. For
$\epsilon_{\rm SF}=10^{-4}$, the limit could be as low as
$P_\alpha/P_{\rm th}<20$. The limits become even more severe for
higher escape fractions.

\section{Contribution from higher order Lyman photons}
\label{app:GHn}

The direct scattering rate of higher order Ly-$n$ photons is
\begin{equation}
P^{\rm dir}_n = P^{\rm inc}_n(0){\cal S}_n,
\label{eq:Pndir}
\end{equation}
where $P^{\rm inc}_n(0) = \sigma_n L_{\nu_n}/(4\pi r_L^2 h\nu_n)$ is
the scattering rate assuming no intergalactic attenuation, $r_L$ is
the luminosity distance from a source with specific luminosity
$L_{\nu_n}$ at the Ly-$n$ transition frequency $\nu_n$, and where
$\sigma_n$ is the total Ly-$n$ resonance cross section. Intergalactic
attenuation is accounted for by the suppression factor ${\cal S}_n$.
To good approximation, ${\cal S}_n\gsim\gamma_n$, where
$\gamma_n=H(z)/[\lambda_n\sigma_nn_{\rm HI}(z)]$ is the Sobolev
parameter for the Ly-$n$ transtion of wavelength $\lambda_n$, $H(z)$
is the Hubble parameter at redshift $z$, and $n_{\rm HI}(z)$ is the
mean intergalactic neutral hydrogen density
\citep{2010MNRAS.402.1780M, 2012MNRAS.426.2129K}. In addition to
direct scatters, higher order Lyman photons will generate lower order
Lyman photons through radiative cascades. The net scattering rate of
higher Lyman photons may be expressed as
\begin{equation} 
P_n = \frac{1}{1-p_{nn}}\sum_{n^\prime=n}^{n_{\rm max}}{\cal  
  C}_{n^\prime n}P^{\rm dir}_{n^\prime},
\label{eq:Pntot}
\end{equation}  
where the scattering cascade matrix is given by
\begin{equation}
{\cal C}_{n^\prime n}=\sum_{n^{\prime\prime}>n}^{n^\prime}{\cal
  C}_{n^\prime n^{\prime\prime}}\eta_{n^{\prime\prime} n},
\label{eq:Cnn}
\end{equation}
where ${\cal C}_{nn}=1$, ${\cal C}_{n^\prime n}=0$ for $n>n^\prime$ and
  $\eta_{n^\prime n}=p_{n^\prime n}/(1-p_{n^\prime n})$, where
  $p_{n^\prime n}$ is the probability for a Ly-$n^\prime$ photon to
  convert into a Ly-$n$ photon on scattering \citep{2010MNRAS.402.1780M,
  2012MNRAS.426.2129K}. For the special case of \Lya\ photons,
\begin{equation}
P_\alpha = P_\alpha^{\rm inc}(0) + {\cal N}_{\rm scatt}\sum_{n^\prime>2}^{n_{\rm max}}{\cal 
  C}_{n^\prime 2}P^{\rm dir}_{n^\prime},
\label{eq:Patot}
\end{equation} 
and ${\cal N}_{\rm scatt}=1/\gamma_S$ is the number of scatters a
\Lya\ photon undergoes in the IGM before redshifting sufficiently far
into the Lorentz wing to escape.

The total heating rate per unit volume due to the scattering of Ly-$n$
photons ($n>2$) at rate $P_n$ per neutral hydrogen atom is
\begin{eqnarray}
G_n = P_n n_{\rm HI} \frac{h\nu_n}{m_a c^2}&\sum_{n^\prime=1}^{n-1}&
\frac{A(n,1;n^\prime,0) + A(n,1;n^\prime,2)}{A_{n,1}}\nonumber\\
&\times&h\nu_{nn^\prime}
\Biggl(1-\frac{T_K}{\langle T_{nn^\prime}\rangle_{\rm H}}\Biggr),
\label{eq:Gn}
\end{eqnarray}
for an IGM kinetic temperature $T_K$ and harmonic mean light temperature
\begin{equation}
\langle T_{nn^\prime}\rangle_{\rm H}=\frac{\nu_{nn^\prime}}{\nu_n}
\frac{\int_0^\infty d\nu\, n_\nu\varphi_V(a_n,\nu)}{\int_0^\infty d\nu\,
\frac{1}{T_n(\nu)}n_\nu\varphi_V(a_n,\nu)},
\label{eq:TnH}
\end{equation}
where
\begin{equation}
T_n(\nu)=-\frac{h}{k}\Biggl(\frac{d\log
  n_\nu}{d\nu}\Biggr)^{-1}
\label{eq:Tnnu}
\end{equation}
for a photon number density $n_\nu$ \citep{2010MNRAS.402.1780M}. Here
$\nu_{nn^\prime}=\nu_L(1/{n^\prime}^2-1/n^2)$ (and $\nu_n=\nu_{n1}$),
where $\nu_L$ is the frequency of the Lyman edge. $A_{n,1}$ is the
total decay rate of the $p$-state with principal quantum number
$n$. (Note $A(n,1;n^\prime,2)$ is undefined for $n^\prime<3$, and
should be regarded as zero.)

A given position in the IGM at redshift $z$ will receive Ly-$n$
photons only from within a restricted redshift range given by
\begin{equation}
1+z_n^{\rm hor} = (1+z)\frac{1-(n+1)^{-2}}{1-n^{-2}},
\label{eq:znhor}
\end{equation}
as photons emitted from greater distances that would have redshifted
to Ly-$n$ will instead be scattered by the Ly-$(n+1)$ transition of a
neutral hydrogen atom along the way \citep{2005ApJ...626....1B}. For a
comoving source luminosity function $\Phi(L)$, representing the number of
sources per unit volume with luminosity between $L$ and $L+dL$, the
total direct scattering rate for Ly-$n$ photons is
\begin{eqnarray}
P_n^{\rm dir, Tot}&=&\int_z^{z_n^{\rm 
  hor}}\,dz\frac{dr_p}{dz}(1+z)^34\pi r_p^2\nonumber\\
&&\times\int_0^\infty\,dL_{\nu_n}\,\frac{\sigma_nL_{\nu_n}}{4\pi
  r_L^2h\nu_n}{\cal S}_n\Phi(L_{\nu_n})\nonumber\\
&\simeq&\int_z^{z_n^{\rm
         hor}}\,dz\frac{dr_p}{dz}(1+z)^3\frac{\sigma_n\epsilon_n}{h\nu_n}{\cal
         S}_n(z),
\label{eq:PndirTot}
\end{eqnarray}
where $\epsilon_n=\int\,dL_{\nu_n}L_{\nu_n}\Phi(L_{\nu_n})$ is the
comoving source emissivity, and the luminosity distance $r_L$ is
approximated as the proper distance $r_p$. A similar expression
applies for $P_\alpha^{\rm dir, Tot}$, with ${\cal S}_2=1$. To good
approximation, $P_n^{\rm dir, Tot}$ may be used in Eq.~(\ref{eq:Gn})
to obtain the total Ly-$n$ heating rate from all sources since
$\langle T_{nn^\prime}\rangle_{\rm H}$ is nearly independent of source
distance \citep{2010MNRAS.402.1780M}, although it is straightforward
to integrate over sources as in Eq.~(\ref{eq:PndirTot}) if preferred,
and is done for the results presented here.

\bigskip  
\section*{acknowledgments}
 
AM acknowledges support from the UK Science and Technology Facilities Council, Consolidated Grant ST/R000972/1. P.M. acknowledges a NASA contract supporting the WFIRST-EXPO Science Investigation Team (15-WFIRST15-0004), administered by GSFC.


\bibliographystyle{mn2e-eprint}
\bibliography{ms}

\begin{thebibliography}{}

\bibitem[\protect\citeauthoryear{{Barkana} \& {Loeb}}{{Barkana} \&
  {Loeb}}{2005}]{2005ApJ...626....1B}
{Barkana} R.,  {Loeb} A.,  2005, \apj, 626, 1

\bibitem[\protect\citeauthoryear{{Blumenthal}, {Faber}, {Primack} \&
  {Rees}}{{Blumenthal} et~al.}{1984}]{1984Natur.311..517B}
{Blumenthal} G.~R.,  {Faber} S.~M.,  {Primack} J.~R.,    {Rees} M.~J.,  1984,
  \nat, 311, 517

\bibitem[\protect\citeauthoryear{{Bond} \& {Szalay}}{{Bond} \&
  {Szalay}}{1983}]{1983ApJ...274..443B}
{Bond} J.~R.,  {Szalay} A.~S.,  1983, \apj, 274, 443

\bibitem[\protect\citeauthoryear{{Bowman}, {Rogers}, {Monsalve}, {Mozdzen} \&
  {Mahesh}}{{Bowman} et~al.}{2018}]{2018Natur.555...67B}
{Bowman} J.~D.,  {Rogers} A. E.~E.,  {Monsalve} R.~A.,  {Mozdzen} T.~J.,
  {Mahesh} N.,  2018, \nat, 555, 67

\bibitem[\protect\citeauthoryear{{Chen} \& {Miralda-Escud{\'e}}}{{Chen} \&
  {Miralda-Escud{\'e}}}{2004}]{2004ApJ...602....1C}
{Chen} X.,  {Miralda-Escud{\'e}} J.,  2004, \apj, 602, 1

\bibitem[\protect\citeauthoryear{{Chuzhoy} \& {Shapiro}}{{Chuzhoy} \&
  {Shapiro}}{2006}]{2006ApJ...651....1C}
{Chuzhoy} L.,  {Shapiro} P.~R.,  2006, \apj, 651, 1

\bibitem[\protect\citeauthoryear{{Couchman} \& {Rees}}{{Couchman} \&
  {Rees}}{1986}]{1986MNRAS.221...53C}
{Couchman} H.~M.~P.,  {Rees} M.~J.,  1986, \mnras, 221, 53

\bibitem[\protect\citeauthoryear{{D'Aloisio}, {McQuinn}, {Trac}, {Cain} \&
  {Mesinger}}{{D'Aloisio} et~al.}{2020}]{2020arXiv200202467D}
{D'Aloisio} A.,  {McQuinn} M.,  {Trac} H.,  {Cain} C.,    {Mesinger} A.,  2020,
  arXiv e-prints, p. arXiv:2002.02467, 2002.02467

\bibitem[\protect\citeauthoryear{{Dijkstra}, {Mesinger} \& {Wyithe}}{{Dijkstra}
  et~al.}{2011}]{2011MNRAS.414.2139D}
{Dijkstra} M.,  {Mesinger} A.,    {Wyithe} J. S.~B.,  2011, \mnras, 414, 2139

\bibitem[\protect\citeauthoryear{{Field}}{{Field}}{1958}]{1958PROCIRE.46..240F}
{Field} G.~B.,  1958, Proc. I.R.E., 46, 240

\bibitem[\protect\citeauthoryear{{Field}}{{Field}}{1959}]{1959ApJ...129..536F}
{Field} G.~B.,  1959, \apj, 129, 536

\bibitem[\protect\citeauthoryear{{Field}}{{Field}}{1964}]{1964ApJ...140.1434F}
{Field} G.~B.,  1964, \apj, 140, 1434

\bibitem[\protect\citeauthoryear{{Haiman}, {Rees} \& {Loeb}}{{Haiman}
  et~al.}{1997}]{1997ApJ...476..458H}
{Haiman} Z.,  {Rees} M.~J.,    {Loeb} A.,  1997, \apj, 476, 458

\bibitem[\protect\citeauthoryear{{Harrington}}{{Harrington}}{1973}]{1973MNRAS.162...43H}
{Harrington} J.~P.,  1973, \mnras, 162, 43

\bibitem[\protect\citeauthoryear{{Heckman}, {Borthakur}, {Overzier},
  {Kauffmann}, {Basu-Zych}, {Leitherer}, {Sembach}, {Martin}, {Rich},
  {Schiminovich} \& {Seibert}}{{Heckman} et~al.}{2011}]{2011ApJ...730....5H}
{Heckman} T.~M.,  {Borthakur} S.,  {Overzier} R.,  {Kauffmann} G.,  {Basu-Zych}
  A.,  {Leitherer} C.,  {Sembach} K.,  {Martin} D.~C.,  {Rich} R.~M.,
  {Schiminovich} D.,    {Seibert} M.,  2011, \apj, 730, 5

\bibitem[\protect\citeauthoryear{{Higgins} \& {Meiksin}}{{Higgins} \&
  {Meiksin}}{2009}]{2009MNRAS.393..949H}
{Higgins} J.,  {Meiksin} A.,  2009, \mnras, 393, 949

\bibitem[\protect\citeauthoryear{{Higgins} \& {Meiksin}}{{Higgins} \&
  {Meiksin}}{2012}]{2012MNRAS.426.2380H}
{Higgins} J.,  {Meiksin} A.,  2012, \mnras, 426, 2380

\bibitem[\protect\citeauthoryear{{Hirata}}{{Hirata}}{2006}]{2006MNRAS.367..259H}
{Hirata} C.~M.,  2006, \mnras, 367, 259

\bibitem[\protect\citeauthoryear{{Kakiichi}, {Meiksin} \& {Tittley}}{{Kakiichi}
  et~al.}{2012}]{2012MNRAS.426.2129K}
{Kakiichi} K.,  {Meiksin} A.,    {Tittley} E.,  2012, \mnras, 426, 2129

\bibitem[\protect\citeauthoryear{{Loeb} \& {Rybicki}}{{Loeb} \&
  {Rybicki}}{1999}]{1999ApJ...524..527L}
{Loeb} A.,  {Rybicki} G.~B.,  1999, \apj, 524, 527

\bibitem[\protect\citeauthoryear{{Madau}}{{Madau}}{2018}]{2018MNRAS.480L..43M}
{Madau} P.,  2018, \mnras, 480, L43

\bibitem[\protect\citeauthoryear{{Madau}, {Meiksin} \& {Rees}}{{Madau}
  et~al.}{1997}]{1997ApJ...475..429M}
{Madau} P.,  {Meiksin} A.,    {Rees} M.~J.,  1997, \apj, 475, 429

\bibitem[\protect\citeauthoryear{{Mason}, {Treu}, {Dijkstra}, {Mesinger},
  {Trenti}, {Pentericci}, {de Barros} \& {Vanzella}}{{Mason}
  et~al.}{2018}]{2018ApJ...856....2M}
{Mason} C.~A.,  {Treu} T.,  {Dijkstra} M.,  {Mesinger} A.,  {Trenti} M.,
  {Pentericci} L.,  {de Barros} S.,    {Vanzella} E.,  2018, \apj, 856, 2

\bibitem[\protect\citeauthoryear{{Meiksin}}{{Meiksin}}{2006}]{2006MNRAS.370.2025M}
{Meiksin} A.,  2006, \mnras, 370, 2025

\bibitem[\protect\citeauthoryear{{Meiksin}}{{Meiksin}}{2010}]{2010MNRAS.402.1780M}
{Meiksin} A.,  2010, \mnras, 402, 1780

\bibitem[\protect\citeauthoryear{{Meiksin}}{{Meiksin}}{2011}]{2011MNRAS.417.1480M}
{Meiksin} A.,  2011, \mnras, 417, 1480

\bibitem[\protect\citeauthoryear{{Ono}, {Ouchi}, {Mobasher}, {Dickinson},
  {Penner}, {Shimasaku}, {Weiner}, {Kartaltepe}, {Nakajima}, {Nayyeri},
  {Stern}, {Kashikawa} \& {Spinrad}}{{Ono} et~al.}{2012}]{2012ApJ...744...83O}
{Ono} Y.,  {Ouchi} M.,  {Mobasher} B.,  {Dickinson} M.,  {Penner} K.,
  {Shimasaku} K.,  {Weiner} B.~J.,  {Kartaltepe} J.~S.,  {Nakajima} K.,
  {Nayyeri} H.,  {Stern} D.,  {Kashikawa} N.,    {Spinrad} H.,  2012, \apj,
  744, 83

\bibitem[\protect\citeauthoryear{{Partridge} \& {Peebles}}{{Partridge} \&
  {Peebles}}{1967}]{1967ApJ...147..868P}
{Partridge} R.~B.,  {Peebles} P.~J.~E.,  1967, \apj, 147, 868

\bibitem[\protect\citeauthoryear{{Planck Collaboration}}{{Planck
  Collaboration}}{2018}]{2018arXiv180706209P}
{Planck Collaboration} 2018, ArXiv e-prints, 1807.06209

\bibitem[\protect\citeauthoryear{{Raiter}, {Schaerer} \& {Fosbury}}{{Raiter}
  et~al.}{2010}]{2010AA...523A..64R}
{Raiter} A.,  {Schaerer} D.,    {Fosbury} R.~A.~E.,  2010, \aap, 523, A64

\bibitem[\protect\citeauthoryear{{Reed}, {Bower}, {Frenk}, {Jenkins} \&
  {Theuns}}{{Reed} et~al.}{2007}]{2007MNRAS.374....2R}
{Reed} D.~S.,  {Bower} R.,  {Frenk} C.~S.,  {Jenkins} A.,    {Theuns} T.,
  2007, \mnras, 374, 2

\bibitem[\protect\citeauthoryear{{Rybicki}}{{Rybicki}}{2006}]{2006ApJ...647..709R}
{Rybicki} G.~B.,  2006, \apj, 647, 709

\bibitem[\protect\citeauthoryear{{Rybicki} \& {dell'Antonio}}{{Rybicki} \&
  {dell'Antonio}}{1994}]{1994ApJ...427..603R}
{Rybicki} G.~B.,  {dell'Antonio} I.~P.,  1994, \apj, 427, 603

\bibitem[\protect\citeauthoryear{{Schauer}, {Glover}, {Klessen} \&
  {Ceverino}}{{Schauer} et~al.}{2019}]{2019MNRAS.484.3510S}
{Schauer} A. T.~P.,  {Glover} S. C.~O.,  {Klessen} R.~S.,    {Ceverino} D.,
  2019, \mnras, 484, 3510

\bibitem[\protect\citeauthoryear{{Schenker}, {Ellis}, {Konidaris} \&
  {Stark}}{{Schenker} et~al.}{2014}]{2014ApJ...795...20S}
{Schenker} M.~A.,  {Ellis} R.~S.,  {Konidaris} N.~P.,    {Stark} D.~P.,  2014,
  \apj, 795, 20

\bibitem[\protect\citeauthoryear{{Seager}, {Sasselov} \& {Scott}}{{Seager}
  et~al.}{2000}]{2000ApJS..128..407S}
{Seager} S.,  {Sasselov} D.~D.,    {Scott} D.,  2000, \apjs, 128, 407

\bibitem[\protect\citeauthoryear{{Shull}, {Harness}, {Trenti} \&
  {Smith}}{{Shull} et~al.}{2012}]{2012ApJ...747..100S}
{Shull} J.~M.,  {Harness} A.,  {Trenti} M.,    {Smith} B.~D.,  2012, \apj, 747,
  100

\bibitem[\protect\citeauthoryear{{Smith}, {Safranek-Shrader}, {Bromm} \&
  {Milosavljevi{\'c}}}{{Smith} et~al.}{2015}]{2015MNRAS.449.4336S}
{Smith} A.,  {Safranek-Shrader} C.,  {Bromm} V.,    {Milosavljevi{\'c}} M.,
  2015, \mnras, 449, 4336

\bibitem[\protect\citeauthoryear{{Smith}, {Wise}, {O'Shea}, {Norman} \&
  {Khochfar}}{{Smith} et~al.}{2015}]{2015MNRAS.452.2822S}
{Smith} B.~D.,  {Wise} J.~H.,  {O'Shea} B.~W.,  {Norman} M.~L.,    {Khochfar}
  S.,  2015, \mnras, 452, 2822

\bibitem[\protect\citeauthoryear{{Stark}, {Ellis}, {Chiu}, {Ouchi} \&
  {Bunker}}{{Stark} et~al.}{2010}]{2010MNRAS.408.1628S}
{Stark} D.~P.,  {Ellis} R.~S.,  {Chiu} K.,  {Ouchi} M.,    {Bunker} A.,  2010,
  \mnras, 408, 1628

\bibitem[\protect\citeauthoryear{{Steidel}, {Erb}, {Shapley}, {Pettini},
  {Reddy}, {Bogosavljevi{\'c}}, {Rudie} \& {Rakic}}{{Steidel}
  et~al.}{2010}]{2010ApJ...717..289S}
{Steidel} C.~C.,  {Erb} D.~K.,  {Shapley} A.~E.,  {Pettini} M.,  {Reddy} N.,
  {Bogosavljevi{\'c}} M.,  {Rudie} G.~C.,    {Rakic} O.,  2010, \apj, 717, 289

\bibitem[\protect\citeauthoryear{{Tinsley}}{{Tinsley}}{1972}]{1972ApJ...178..319T}
{Tinsley} B.~M.,  1972, \apj, 178, 319

\bibitem[\protect\citeauthoryear{{Truran} \& {Cameron}}{{Truran} \&
  {Cameron}}{1971}]{1971ApSS..14..179T}
{Truran} J.~W.,  {Cameron} A.~G.~W.,  1971, \apss, 14, 179

\bibitem[\protect\citeauthoryear{{Tseliakhovich} \& {Hirata}}{{Tseliakhovich}
  \& {Hirata}}{2010}]{2010PhRvD..82h3520T}
{Tseliakhovich} D.,  {Hirata} C.,  2010, \prd, 82, 083520

\bibitem[\protect\citeauthoryear{{Verhamme}, {Dubois}, {Blaizot}, {Garel},
  {Bacon}, {Devriendt}, {Guiderdoni} \& {Slyz}}{{Verhamme}
  et~al.}{2012}]{2012AA...546A.111V}
{Verhamme} A.,  {Dubois} Y.,  {Blaizot} J.,  {Garel} T.,  {Bacon} R.,
  {Devriendt} J.,  {Guiderdoni} B.,    {Slyz} A.,  2012, \aap, 546, A111

\bibitem[\protect\citeauthoryear{{Visbal}, {Barkana}, {Fialkov},
  {Tseliakhovich} \& {Hirata}}{{Visbal} et~al.}{2012}]{2012Natur.487...70V}
{Visbal} E.,  {Barkana} R.,  {Fialkov} A.,  {Tseliakhovich} D.,    {Hirata}
  C.~M.,  2012, \nat, 487, 70

\bibitem[\protect\citeauthoryear{{Whalen}, {van Veelen}, {O'Shea} \&
  {Norman}}{{Whalen} et~al.}{2008}]{2008ApJ...682...49W}
{Whalen} D.,  {van Veelen} B.,  {O'Shea} B.~W.,    {Norman} M.~L.,  2008, \apj,
  682, 49

\bibitem[\protect\citeauthoryear{{Whalen}, {Johnson}, {Smidt}, {Meiksin},
  {Heger}, {Even} \& {Fryer}}{{Whalen} et~al.}{2013}]{2013ApJ...774...64W}
{Whalen} D.~J.,  {Johnson} J.~L.,  {Smidt} J.,  {Meiksin} A.,  {Heger} A.,
  {Even} W.,    {Fryer} C.~L.,  2013, \apj, 774, 64

\bibitem[\protect\citeauthoryear{{Wouthuysen}}{{Wouthuysen}}{1952}]{1952AJ.....57R..31W}
{Wouthuysen} S.~A.,  1952, \aj, 57, 31

\bibitem[\protect\citeauthoryear{{Yoshida}, {Abel}, {Hernquist} \&
  {Sugiyama}}{{Yoshida} et~al.}{2003}]{2003ApJ...592..645Y}
{Yoshida} N.,  {Abel} T.,  {Hernquist} L.,    {Sugiyama} N.,  2003, \apj, 592,
  645

\bibitem[\protect\citeauthoryear{{Zanstra}}{{Zanstra}}{1949}]{1949BAN....11....1Z}
{Zanstra} H.,  1949, \bain, 11, 1

\end{thebibliography}

\label{lastpage}

\end{document}